\documentclass[twocolumn]{aastex631}

\usepackage{natbib}
\usepackage{amsmath}
\usepackage{bm}
\usepackage{soul}

\shorttitle{3-D simulations of massive main-sequence stars III}
\shortauthors{Mao et al.}

\graphicspath{{./}{figs/}}

\newcommand{\code}[1]{\texttt{#1}}

\newcommand{\paperone}{Paper~I} 
\newcommand{\papertwo}{Paper~II} 

\newcommand{\lSect}[1]{{\label{sec:#1}}}
\newcommand{\lFig}[1]{{\label{fig:#1}}}
\newcommand{\lEq}[1]{{\label{eq:#1}}}
\newcommand{\lTab}[1]{{\label{tab:#1}}}
\newcommand{\Tab}[1]{{Table~\ref{tab:#1}}}
\newcommand{\Fig}[1]{{Fig.~\ref{fig:#1}}}

\newcommand{\Eq}[1]{{Eq.~\ref{eq:#1}}}
\newcommand{\Sect}[1]{{\S\ref{sec:#1}}}

\newcommand{\unitstyle}[1]{\ensuremath{\mathrm{#1}}}
\newcommand{\Msun}{\ensuremath{\unitstyle{M}_\odot}}
\newcommand{\natlog}[2]{\ensuremath{#1\times 10^{#2}}}
\newcommand{\hour}{\unitstyle{h}}
\newcommand{\Mm}{\unitstyle{Mm}}
\newcommand{\Hz}{\unitstyle{Hz}} 
\newcommand{\yr}{\unitstyle{yr}}
\newcommand{\cm}{\unitstyle{cm}}
\newcommand{\minute}{\unitstyle{min}}
\newcommand{\numberspace}{\ensuremath{\;}}
\newcommand{\unit}[2]{\ensuremath{#1\numberspace\mathrm{#2}}}

\usepackage[]{color}

\begin{document}

\title{3-D hydrodynamic simulations of massive main-sequence stars. III. The effect of radiation pressure and diffusion leading to a 1-D equilibrium model}
\author{Huaqing Mao}
\affiliation{LCSE and Department of Physics and Astronomy, University of Minnesota, Minneapolis, MN 55455, USA}
\affiliation{Joint Institute for Nuclear Astrophysics - Center for the Evolution of the Elements, USA}

\author{Paul Woodward}
\affiliation{LCSE and Department of Physics and Astronomy, University of Minnesota, Minneapolis, MN 55455, USA}
\affiliation{Joint Institute for Nuclear Astrophysics - Center for the Evolution of the Elements, USA}

\author{Falk Herwig}
\affiliation{Astronomy Research Centre and Department of Physics and Astronomy, University of Victoria, Victoria, BC, V8W 2Y2, Canada}
\affiliation{Joint Institute for Nuclear Astrophysics - Center for the Evolution of the Elements, USA}

\author{Pavel A. Denissenkov}
\affiliation{Astronomy Research Centre and Department of Physics and Astronomy, University of Victoria, Victoria, BC, V8W 2Y2, Canada}
\affiliation{Joint Institute for Nuclear Astrophysics - Center for the Evolution of the Elements, USA}

\author{Simon Blouin}
\affiliation{Astronomy Research Centre and Department of Physics and Astronomy, University of Victoria, Victoria, BC, V8W 2Y2, Canada}
\affiliation{Joint Institute for Nuclear Astrophysics - Center for the Evolution of the Elements, USA}

\author{William Thompson}
\affiliation{Astronomy Research Centre and Department of Physics and Astronomy, University of Victoria, Victoria, BC, V8W 2Y2, Canada}

\author{Benjamin McDermott}
\affiliation{LCSE and Department of Physics and Astronomy, University of Minnesota, Minneapolis, MN 55455, USA}

\begin{abstract}
We present 3-D hydrodynamical simulations of core convection with a stably stratified envelope
of a \unit{25}{\Msun} star in the
early phase of the main-sequence. We use the explicit gas-dynamics code \code{PPMstar} which tracks
two fluids and includes radiation pressure and radiative diffusion. Multiple series of simulations with different luminosities and radiative thermal conductivities are presented. The entrainment rate at the convective boundary, internal gravity waves in and above the boundary region, and the approach to dynamical equilibrium shortly after a few convective turnovers are investigated. We perform very long simulations on $896^3$ grids accelerated by luminosity boost factors $1000$, $3162$ and $10000$. In these simulations the growing penetrative convection reduces the initially unrealistically large entrainment.  This reduction is enabled by a spatial separation that develops between the entropy gradient and the composition gradient. The convective boundary moves outward much more slowly at the end of these simulations. Finally, we present a 1-D method to predict the extent and character of penetrative convection beyond the Schwarzschild bounxdary. The 1-D model is based on a spherically-averaged reduced entropy equation that takes the turbulent dissipation as input from the 3-D hydrodynamic simulation and takes buoyancy and all other energy sources and sinks into account. This 1-D method is intended to be ultimately deployed in 1-D stellar evolution calculations and is based on the properties of penetrative convection in our simulations carried forward through the local thermal timescale.
\end{abstract}


\section{Introduction}
Convective transport can be very efficient in stellar interiors, owing to the high energy densities there \citep{kippenhahn1990stellar}. At the convective-radiative boundary, it can play a crucial role in mixing chemical species \citep[e.g.][ in novae]{Denissenkov:2012cu}.
Yet convection is one major uncertainty in the 1-D stellar evolution model \citep[e.g.][ in massive stars]{Sukhbold:2014,Davis:2018jz,Kaiser:2020}, with a set of parameters to calibrate to match with the observations \citep[e.g.][]{Schaller:1992vq,Ribas:2000fn,Trampedach2014,Tkachenko2020a, higl2021calibrating}.
For example, the efficiency of convective boundary mixing (CBM) during the main-sequence directly affects the model's brightness and main-sequence lifetime \citep{salaris2017,Higgins:2019}.
The local theory of convection, mixing-length theory (MLT) formalized by \cite{Bohm1958} and \cite{cox1968principles} is widely used in 1-D stellar evolution codes \citep[e.g.][]{Paxton:2010}. Other sophisticated theories on convection have also been proposed. For example, \cite{xiong1986} developed a non-local MLT that indicates penetrative convection. \cite{pasetto2014theory} removes the mixing length in their convection theory. A spectrum of turbulent eddies instead of a typical rising blob is considered in \cite{canuto1991stellar}.

Convection is not only an important mechanism to transport energy and species, but also excites internal gravity waves (IGWs) \citep{lecoanet2013internal, pinccon2016generation}.
It is predicted theoretically that radiative diffusion damps travelling IGWs, which carry angular momentum \citep{rogers:17,Aerts:2019ARAA}.
This process leads to deposition of angular momentum where the IGWs are damped, and hence to redistribution of angular momentum \citep{Zahn:97}.
Asteroseismological observations help constrain convective boundary mixing and diffusive mixing in the radiative envelope \citep{Moravveji:2015kr,Michielsen:2019ht,michielsen:21}.

Penetrative convection has been investigated in theory and through numerical simulations for decades in various contexts, core convections and shell convections for exapmle \citep{roxburgh1989integral,arnett2015beyond,anders:21,
korre2021dynamics,Blouin_RGB:23b}. The extent of convective penetration and its dependence on various properties of the Schwarzschild boundary (SB) have been studied \citep{hurlburt1994penetration,Baraffe2021}. The temperature gradient in the convective boundary (CB) region may be deduced by asteroseismological observation and modeling \citep{michielsen:21}. Current treatment of the convective boudary in 1-D stellar evolution simulations includes f overshooting \citep{Herwig:2000ua}, instantaneous overshooting \citep{Maeder:1976} and entrainment \citep{staritsin2013turbulent,scott2021convective}.
In this work, we define the SB to be the location where the rising radiation diffusion energy flux as we go outward in radius in the core convection zone first equals the total luminosity.  We find that this is not the location where the entropy gradient first becomes positive and the temperature gradient first becomes subadiabatic, as we will discuss later.  Beyond the SB we have a region of penetrative convection leading up to the CB.  We here define the CB to be that radius at which the radiative energy flux becomes equal to the total luminosity, the convective entropy flux vanishes, and also the turbulent dissipation of kinetic energy of the convection flow vanishes.

Previously, in the first paper of this series, we have introduced the general properties of core-convection simulations of a
25 \Msun\ star approximated with an ideal gas equation of state \citep[\paperone]{herwig:23a}.
We confirmed earlier results of massive main-sequence star simulations by \cite{Meakin:2007dj},
\cite{Gilet:2013bj} { and more recently by  \cite{Baraffe:23} that} entrainment rates of envelope material into the convective core
are orders of magnitude larger than what is compatible with stellar models and basic observational properties.
{ These large entrainment rates are the response of the 3-D hydrodynamic simulation to a radial stratification, for example from a 1-D initial state, that is not in dynamic and thermal equilibrium.}

The properties of IGWs in our 3-D \texttt{PPMstar} ideal gas simulations are presented in \citet[][\papertwo]{Thompson:2023a}. One important aspect
of IGWs excited by core convection is the possibility that they may cause material or angular momentum mixing in the radiative layer. Radiative diffusion permits
the entropy in the stably stratified envelope to no longer be a constant of the motion. As a consequence, irreversible envelope mixing becomes possible,
even though IGW velocity amplitudes are damped by radiative diffusion. Our strategy in this paper is to study the impact of radiation pressure and
radiative diffusion on the convection zone in our model star and on the structure of the CB region. We will analyze the spectrum
of IGWs that are excited at the CB for the purpose of comparison with the studies of \paperone\ and \papertwo\ in this series, but we will leave the
issue of potential material mixing in the envelope to a forthcoming paper.

The main goals of this work are as follows: to test whether adopting a more realistic simulation approach which includes radiation pressure and diffusion can reduce the entrainment rate significantly; to study the effect of radiative diffusion on the spectrum of IGWs in the stable envelope; to investigate the stratification of penetrative convection and develop a method to predict the convective penetration depth.

The first 3 sections discuss flow phenomena on a short timescale (convective timescale) and the following two sections investigate the growing penetrative convection on a thermal timescale.
Finally, we discuss our results and conclusions in the last section.
Specifically, in \Sect{methods} we present the simulation method, simulation setup, and assumptions. 
Section \Sect{quasi-steady} describes the general flow dynamics from the onset of core convection to a 3-D quasi-steady state on a convective timescale, introduces CBM, excitation of IGWs and their power spectra, and discusses the effect of radiative diffusion on CBM and IGWs.
Simulations of different luminosities, thermal conductivities, and resolutions are tabulated in \Tab{run_tab}, with their entrainment rates that quantify the efficiency of CBM.
In \Sect{long-term}, the long-time behaviors of stellar stratification and convective penetration are discussed.
The gradual development of the penetration region beyond the SB is observed in a very long duration simulation.  In this simulation the development of a positive entropy gradient in the penetration region that is sustained despite efficient species mixing is identified as a key structure that acts to bring the intensity of convective motions down, so that further entrainment and outward motion of the convective boundary is greatly reduced.
In \Sect{penetration}, a method to predict the penetration depth and the stratification within the penetration region is presented in terms of a 1-D model of the core convection zone that can be worked out if the kinetic energy dissipation rate up to the SB has either been determined from a short 3-D simulation on a modest grid or has been approximated by interpolating between such simulations under similar conditions.
We summarize and discuss our main results and conclusions in \Sect{conclusions}.
 
\section{Methods and assumptions}
\lSect{methods}

To study the effect of radiation, we apply the equation of state that includes radiation pressure in
addition to that of a monatomic gas. This allows direct application of the \code{MESA} \citep{Paxton:2010,Paxton:2013,Paxton:2015} model
 with minimal fitting and approximation in going from 1-D to 3-D initialization. The base state is constructed from the 25 \Msun \code{MESA} stellar evolution model \citep{Davis:2018jz} \natlog{1.64}{6}\yr\ after the start of H burning on the zero-age main sequence. The exponential CBM model is used. In this model, the region outside the SB obeys the radiative temperature gradient. Details on the 1-D model can be found in \paperone. \Fig{base-MESA-compare} shows the agreement of radial profiles of the initial state on the 3-D Cartesian grid with the \code{MESA} model.

\begin{figure}
% Notebook: svn repo:
% ppmstar/projects/H-core-M25/setup/setup_H-core_rad.ipynb
%/user/scratch14_ppmstar-lcse/fherwig/ppmstar/projects/H-core-M25/setup
%  \includegraphics[width=\columnwidth]{figs/R-mass-basestate-MESA.pdf}
%  \includegraphics[width=\columnwidth]{figs/R-P-basestate-MESA.pdf}
  \includegraphics[width=\columnwidth]{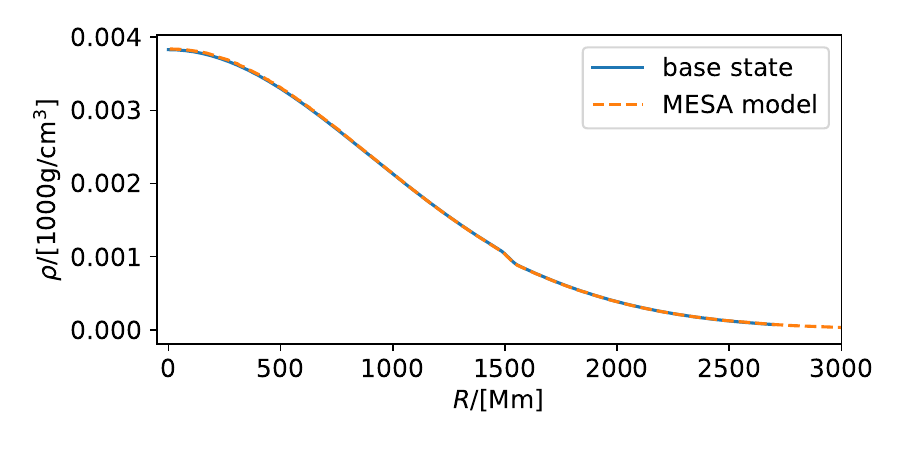}
  \includegraphics[width=\columnwidth]{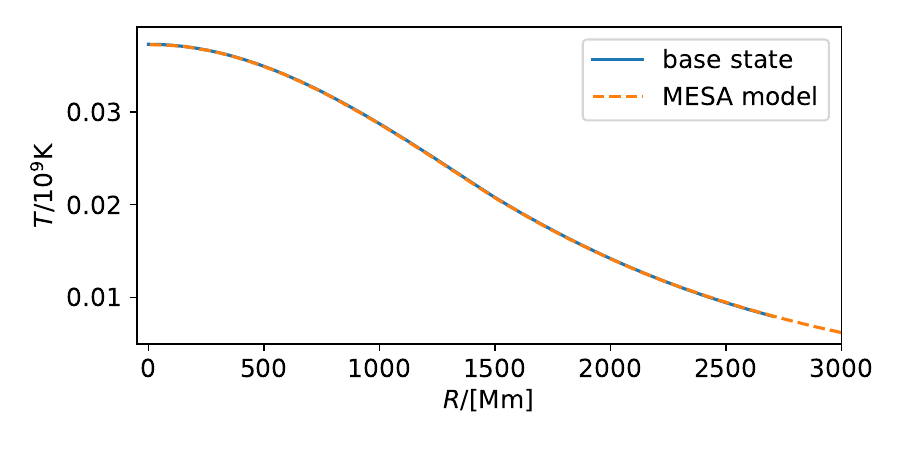}
  \caption{Comparison of adopted base state for the 3-D simulations and
    the \code{MESA} radial profile of density and temperature. Quantities are given in their code units. }
  \lFig{base-MESA-compare}
\end{figure}

We use the \code{PPMstar} gas dynamics code described in \cite{Woodward:2013uf} and applied in
\cite{Woodward:2013uf,Jones:2017kc,Andrassy:2020}. The \code{PPMstar} code tracks the H-rich materials in the stable envelope by fractional volume $f_{\rm V}$, and materials in the convective core by $1 - f_{\rm V}$. The mean molecular weight of each cell is a weighted average of the mean molecular weights of the envelope material and the core material,

\begin{eqnarray}
 \mu = f_\mathrm{V}\cdot\mu_\mathrm{env}+(1-f_\mathrm{V})\cdot\mu_\mathrm{core}. \lEq{mu}
\end{eqnarray}

Here, $\mu_\mathrm{env}=0.6171$ and $\mu_\mathrm{core}=0.6689$. The simulations are initialized such that the $f_\mathrm{V}$ transitions from 0 to 1 at the convective boundary, as $\mu$ changes from $\mu_\mathrm{core}$ to $\mu_\mathrm{env}$.
In this version, the contribution of radiation is included in the internal energy per unit mass $e$, pressure $p$ and specific entropy $s$ according to the equation of state relations:

\begin{eqnarray}
        e(\rho,T,\mu)&=&\frac{RT}{(\gamma-1)\mu}+\frac{a T^4}{\rho}\\
        p(\rho,T,\mu)&=&\frac{R\rho T}{\mu}+\frac{a T^4}{3} \\
        s(\rho,T,\mu)&=&-\frac{R}{\mu}\ln \rho+\frac{R}{(\gamma-1)\mu}\ln T+\frac{4aT^3}{3\rho},
\end{eqnarray}

where $\mu$ is computed by \Eq{mu}, $\rho$ is the density, $T$ is the temperature, $R$ is the gas constant, $\gamma=5/3$, and $a$ is the radiation constant.
In the \code{PPMstar}\ code, a model equation of state \citep{woodward:86} is fitted to local conditions in each grid cell and upon each time step:

\begin{eqnarray}
   p = p_{00}+(\tilde{\gamma}-1)\rho e
\end{eqnarray}

Here the coefficients $p_{00}$ and $\tilde{\gamma}$ are determined upon the outset of every time step in each grid cell such that the correct sound speed $c_s$ and energy density $\rho e$ are recovered:

\begin{eqnarray}
   \tilde{\gamma}=1+\frac{c_s^2\rho}{p+\rho e}, \
   p_{00}=p-(\tilde{\gamma}-1)\rho e.
\end{eqnarray}

The radiative flux,

\begin{eqnarray}
    \bm{F}=-k\nabla T
\end{eqnarray}

is implemented explicitly in \code{PPMstar} as a part of the energy flux in every time step update, with radiative thermal conductivity \citep{kippenhahn1990stellar}
\begin{eqnarray}
    k = \frac{4acT^3}{3\kappa\rho}.
\end{eqnarray}

Here, $\kappa$ is the opacity, $c$ is the speed of light.
Specifically, the interface values are taken for $\kappa$ and $\rho$ and the temperature gradient is calculated by differencing the cell averages of temperature from the grid cells on the left of the interface and of the right of the interface.
Simulations from M200 to M213 (see \Tab{run_tab}) use the following opacity fit as a function of hydrogen mass fraction $x_\mathrm{H}$ and temperature:

\begin{eqnarray}
     \kappa &=& \min(\frac{c_{es}10^{\sum_{i=0}^{3} (a_i(\log_{10}T)^{3-i})}}{\kappa_{\rm min}c_{\rm corr}}, \kappa_{\rm tot} )\\
     c_{\rm es} &=& 0.2(1+x_{\rm H}) \nonumber \\
     c_{\rm corr} &=& 1+0.5( \frac{\kappa_{\rm max}}{\kappa_{\rm min}}-1)(1-\tanh(w\log_{10}\frac{T}{T_0})). \nonumber
\end{eqnarray}
The coefficients above have the following values:

\begin{linenomath}
\small{
\[
\begin{array}{ccc}
  \kappa_{\rm min}, \kappa_{\rm max} =  \begin{bmatrix}
  0.32009677883170023& 0.3420244849271527   \\
    \end{bmatrix},
\end{array}
\]
}

$\kappa_{tot}=0.66$, $\mathrm{log}_{10}T_0=7.06$, $w=13.0$, and

\small{
\[
\begin{array}{ccc}
  \mathbf{a} =  \begin{bmatrix}
  -0.21193353 & 4.58822546& -33.25338915& 80.22027956   \\
    \end{bmatrix}.
\end{array}
\]
}
\end{linenomath}

Simulations M284, M250, M251 and M252 use another opacity fit to the OPAL opacity \citep{iglesias1996updated} as a function of density, temperature and hydrogen mass fraction:

\begin{eqnarray}
     \kappa = \sum_{i=0}^{5} a_i (t_7)^{5-i}
     \lEq{kappa} 
\end{eqnarray}
In \Eq{kappa}, $t_7 = \log_{10}T - 7 $ and
\begin{eqnarray}
     & a_i = w_{11}a^i_{11}+w_{12}a^i_{12} + w_{21}a^i_{21}+w_{22}a^i_{22},   \nonumber \\
     & w_{11} = (r_2-r)(x_2-x_\mathrm{H})/((r_2-r_1)(x_2-x_1)),   \nonumber \\
     & w_{12} = (r_2-r)(x_\mathrm{H}-x_1)/((r_2-r_1)(x_2-x_1)),   \nonumber \\
     & w_{21} = (r-r_1)(x_2-x_\mathrm{H})/((r_2-r_1)(x_2-x_1)),  \nonumber \\
     & w_{22} = (r-r_1)(x_\mathrm{H}-x_1)/((r_2-r_1)(x_2-x_1)),  \nonumber \\
     & r = \log_{10}\rho -3\log_{10}T + 21  \nonumber
\end{eqnarray}

where

\small{
\[
\begin{array}{ccc}
    \begin{bmatrix}
 \mathbf{a}_{11}  & \mathbf{a}_{12}  & \mathbf{a}_{21}  & \mathbf{a}_{22}  \\
    \end{bmatrix} &
    = &
\end{array}
\]
}

\small{
\[
\begin{bmatrix}
    -3.517557& -3.80745443& -2.89317784& -3.30952768 \\
    5.08986892& 5.62926873& 5.70900505& 6.48609904 \\
    -2.26446972& -2.57639712& -3.44350707& -3.91848895\\
    0.18902794& 0.23571381& 0.57976535& 0.67356251\\
    -0.01369687& 0.01629371& -0.06750886& -0.07837157\\
    0.33163233& 0.37333938& 0.35946486& 0.40405573
\end{bmatrix},
\]
}

\small{
\[
\mathbf{x} = \begin{bmatrix}
       0.5562991483419806 & 0.7564365605920813
     \end{bmatrix}^\mathrm{T},
\]
}

\small{
\[
\mathbf{r} = \begin{bmatrix}
     -4.178136499095293& -3.7948565951463475
     \end{bmatrix}^\mathrm{T}.
\]
}

The resulting opacities are in cgs units.

We apply a reflecting boundary condition at radius 2670 \Mm, and make the heat fluxes at opposite cell interfaces equal for
3 grid cell widths inside this reflecting sphere.
We perform a series of $25\Msun$ simulations (\Tab{run_tab}), with varying
driving luminosities and radiative thermal conductivity $k$. Properties such as the mass
entrainment rate at the CB at the nominal luminosity are extrapolated from
simulations with boosted luminosities.
For a luminosity boosting factor $X$, we have cases with 0, $X^{2/3}$ and $X$ boosting factors for radiative diffusion. Henceforth, we refer to them by no diffusion, intermediate diffusion, and high diffusion.

\section{From the initial transient to a quasi-steady 3-D flow}
\lSect{quasi-steady}

Here we briefly describe the dynamics of the initial transient and the following quasi-steady 3-D flow. The initial transient is complete after the first few convective turn-over times for the largest eddies. In our many cases considered here, we find that the visualization looks qualitatively similar regardless of the boosting factor for luminosity and radiative diffusion. See our representative simulation M252 (luminosity and radiative diffusion boosted by a factor of 10000) at \href{https://ppmstar.org/}{https://ppmstar.org} as well as at \href{https://www.lcse.umn.edu} {https://www.lcse.umn.edu}. In the discussion below, we will point out the effect of radiative diffusion when it matters qualitatively and quantitatively.

\subsection{The development of the fully convective core}
At time 0, the initial state is in perfect hydrostatic equilibrium. The radiative diffusion is transporting heat according to the stratification and opacity. As in \paperone\, the nuclear burning is emulated as a time-independent Gaussian volume heating $\sim \exp(-r^2/(2 \sigma^2))$, $\sigma=280\,\Mm$. The change of chemical composition due to nuclear burning is negligible on the timescale that we simulate and thus ignored as an approximation. Given the temperature gradient, there is the excess heat in the core accumulating due to insufficient radiative energy transport. The center of the core becomes convectively unstable as a result. The central gas parcels rise because of the buoyancy force and thereby convection starts. Because the convective core is almost adiabatic, the moving fluid elements move effortlessly on the same adiabat. The excess heat unable to be carried by the radiative diffusion is now transported by the emerging convection within the core until the rising, relatively buoyant fluid elements encounter the positive entropy gradient where the stratification becomes convectively stable.

Once the rising plumes encounter the entropy gradient, the buoyancy force restrains them from going further outward in radius. The interaction between the plumes and the convective-radiative boundary excites IGWs that propagate in the stable envelope. During the first few convective turnovers, the core convection becomes fully turbulent and excites IGWs of a broad range of wavelengths. An analysis of the power spectrum of the IGWs in the stable envelope after the initial transient adjustment of the flow to its 3-D degrees of freedom is presented at the end of this section.

\begin{figure*}
       \includegraphics[width=\textwidth]{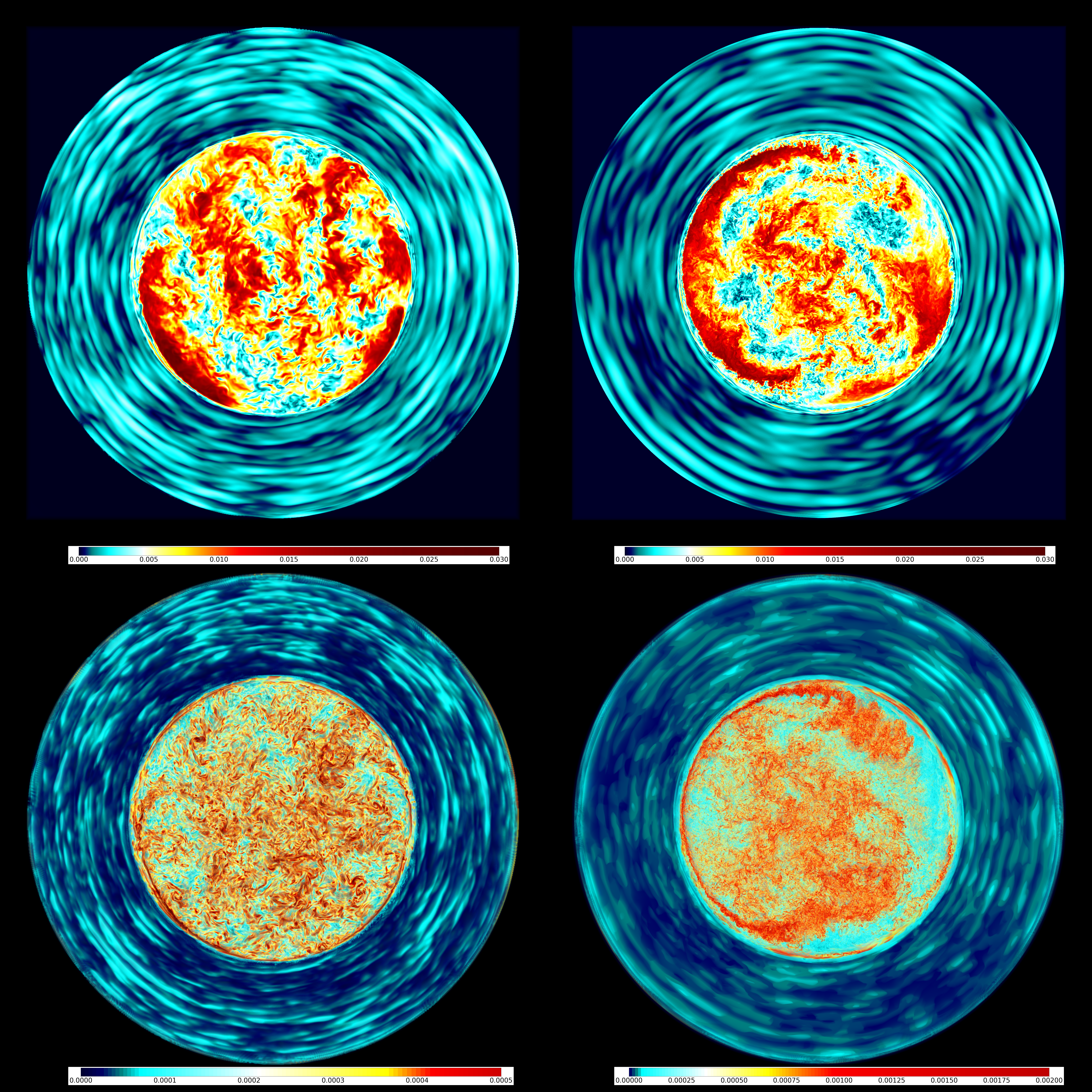}
       \caption{Images of a thin slice through the center of the star of the horizontal velocity component $u_\mathrm{h}$ (top row) of M201 (left column, no radiative diffusion, $1152^3$ grid, time 3183.2 hrs) and M284 (right column, 1000x radiative diffusion, $2688^3$ cells, time 3266 hrs), and of the vorticity magnitude $|\nabla \times \mathbf{u}|$ (bottom row).  The units used in the colorbars are $\mathrm{Mm}\ s^{-1}$ on the top row and $s^{-1}$ on the bottom row. Movies of these quantities are available at \href{https://ppmstar.org}{https://ppmstar.org} as well as at \href{https://www.lcse.umn.edu} {https://www.lcse.umn.edu}. In the images on the top row, internal gravity wave (IGW) motions excited by the convection are clearly visible in the elongated blue and aqua-white features that delineate the mostly horizontal gas motions in these waves. In the vorticity images on the bottom row, the shear from these wave motions in the stably stratified envelope, shown again in blue and aqua-white, has amplitudes an order of magnitude or more smaller than the vorticity values in the turbulent convection zones of these two simulations. The length scale in these images can be determined by the radius of the convective boundary, which is 1530 Mm in the images on the right and ia 1546 Mm in the images on the left.}
    \lFig{image}
\end{figure*}
\begin{table}
        \centering
        \caption{Simulation summary  
          providing the run ID, the grid,
          luminosity $L$ boosting factor, thermal
          conductivity $k$ boosting factor, end time of the run, and entrainment rate, $*$ denotes values from the \code{MESA} model. The runs labelled by $\dagger$ are long-duration, the entrainment rates of which decline over time. Hence we fit them by a straight line from  14323 to 17188 hours to compute the corresponding entrainment rates.}
        \lTab{run_tab}
        \begin{tabular}{lrrrrr}
                \hline
%                ID & grid &  $L$ &$K$ & $t_\mathrm{end}/h$ &$\dot M_1/[\Msun/\second]$\\
                ID & grid &  $L/L_*$ &$k/k_*$ & $t_\mathrm{end}/{\rm h}$ &$\dot M/[\Msun\ \yr^{-1}]$\\
                \hline
                                M200&   $768^3$ & 1000.0 & 0.0 & 1817.6& \natlog{6.82}{-1}\\ % \natlog{2.16}{-8}
                                M201&  $1152^3$ & 1000.0 & 0.0 & 3556.3& \natlog{6.85}{-1}\\ % \natlog{2.17}{-8}
                                M202&  $1152^3$ &  100.0 & 0.0 & 2439.2& \natlog{3.60}{-2}\\ % \natlog{1.14}{-9}
                                M203&  $1152^3$ & 3162.0 & 0.0 & 1468.1& \natlog{2.41}{0}\\ % \natlog{7.65}{-8}
                                M204&  $1152^3$ & 1000.0 & 100.0 & 3362.9& \natlog{6.53}{-1}\\ % \natlog{2.07}{-8}
                                M205&  $1152^3$ &  100.0 & 21.5 & 2648.4& \natlog{3.91}{-2}\\ % \natlog{1.24}{-9}
                                M206&  $1152^3$ & 3162.0 & 215.4 & 1549.8& \natlog{2.16}{0}\\ % \natlog{6.85}{-8}
                                M207&  $1152^3$ & 1000.0 & 1000.0 & 3838.4& \natlog{3.69}{-1}\\ % \natlog{1.17}{-8}
                                M208&  $1152^3$ &  100.0 & 100.0 & 2446.4& \natlog{2.00}{-2}\\ % \natlog{6.34}{-10}
                                M209&  $1152^3$ & 3162.3 & 3162.3 & 1465.3& \natlog{1.36}{0}\\ % \natlog{4.32}{-8}
                                M210&  $1728^3$ & 1000.0 & 1000.0 & 3495.3& \natlog{3.91}{-1}\\ % \natlog{1.24}{-8}
                                M211&   $768^3$ & 1000.0 & 100.0 & 2089.7& \natlog{6.31}{-1}\\ % \natlog{2.00}{-8}
                                M212&  $1152^3$ &  31.62 & 31.62 & 2297.4& \natlog{6.03}{-3}\\ % \natlog{1.91}{-10}
                                M213&   $768^3$ & 1000.0 & 1000.0 & 3537.5& \natlog{3.72}{-1}\\  % \natlog{1.18}{-8}
                                M284&  $2688^3$ & 1000.0 & 1000.0 & 4669.3& \natlog{3.38}{-1}\\ % \natlog{1.07}{-8}
                                M250$^\dagger$&  $896^3$ &  3162.3 &  3162.3 & 20769.0& \natlog{5.77}{-1} \\ % \natlog{1.83}{-8} \natlog{2.25}{-8}
                                M251$^\dagger$&  $896^3$ &  1000.0 &  1000.0 & 18444.4& \natlog{1.74}{-1} \\ % \natlog{5.51}{-9} \natlog{7.09}{-9}
                                M252$^\dagger$&  $896^3$ & 10000.0 & 10000.0 & 25137.6& \natlog{1.40}{-1} \\ % \natlog{4.45}{-9} \natlog{3.45}{-8}
                                \hline
        \end{tabular}
\end{table}
\begin{figure}
       \includegraphics[width=\columnwidth]{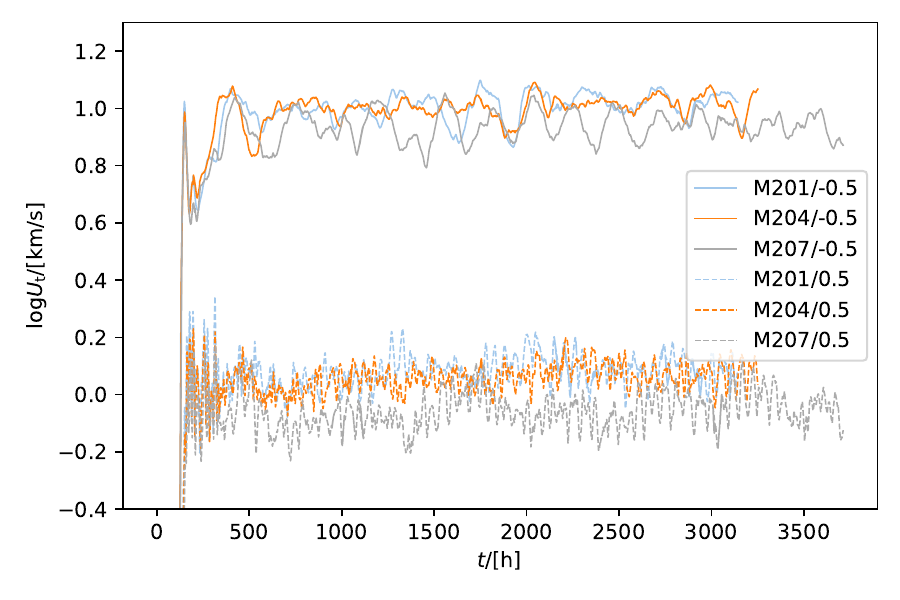}
       \caption{The overall magnitude of horizontal velocities 0.5 $H_p$ below and above the $N^2$ peak (see \Eq{BV} ) becomes constant after an initial transient (400 hours) when we average over the persistent fluctuations.  The 3 runs shown all have grids of $1152^3$ cells.  M201, M204, and M207 have no, intermediate, and high diffusion, respectively. }
    \lFig{Ut-vs-time}
% Falk provides this plot.
\end{figure}

The convective core soon develops the characteristic dipole circulation pattern that was first seen in the 3-D simulations of \cite{Porter:2000kv}. It has been noted by many investigators that convection tends to develop convection cells that extend to the largest vertical scale \citep{hurlburt:86, Freytag:vw, Porter:2000kv, andrassy:22}. In \Fig{image}, when the dipole plume hits the CB and diverges, the flows become mostly horizontal near the boundary, bringing along buoyant materials from the boundary.  This behavior is evident in both images at the top in \Fig{image} from the red lanes of very high horizontal velocity $u_\mathrm{h}$ perpendicular to the radial direction, which are seen along the CB in both images. Entrainment of the fluid from the stable layer into the convection zone is facilitated by the boundary layer separation, as discussed in \cite{Woodward:2013uf}. This boundary layer separation occurs when the flows along the boundary collide and are forced downward toward the center of the star, bringing some of the entrained gas from above the CB with them. In the lower images in \Fig{image}, the highest vorticities shown in yellow and red delineate the strong shear layer where the gas of the convection zone flows along the CB and later separates from it. This shear layer is more difficult to identify in the image from run M201 at the bottom-left, because this run with no diffusion cannot generate a region of penetrative convection.  In the image from run M201, a movie (at \href{https://ppmstar.org}{https://ppmstar.org} as well as at \href{https://www.lcse.umn.edu} {https://www.lcse.umn.edu}) makes clear that the upwelling of the global dipole circulation is aimed roughly at 5:30 o'clock, and the flows along the boundary separate at roughly 9 o'clock and 2 o'clock.  It is hard to trace the shear layers in this image, because they are pressed right up against the CB. In the configuration seen in the image from run M284 at the right, the position of the shear layer shows that it is separated from the CB by a thin layer of gas along most of its length. This is a signature of penetrative convection. The cause for this difference in behavior is discussed in \Sect{long-term}. Simply stated, in the absence of heat transport by radiation diffusion in run M201, heat energy is being transported outward by convection right up to the CB. This heat cannot be transported further outward in M201, because the convection stops at the CB. Hence heat must accumulate inside the CB, and as a whole the convection zone must therefore slowly expand. In run M284, with high diffusion, the radiation transports heat outward at more than the full luminosity in a significant region of convective penetration between the SB, at roughly 1420 Mm, and the CB, at 1530 Mm. Radiation then carries the full luminosity outward beyond the CB.  In the penetration region inside the CB, convective heat transport is inward rather than outward, and the turbulence of the convective flow is less vigorous.  We will see in \Sect{long-term} how this all works out in detail.

In both flows shown in \Fig{image}, a state of dynamical equilibrium is achieved in the relatively short time of several turn-over times of the largest convective eddies, that is, of the large dipole circulation. We define dynamical equilibrium as a state in which the kinetic motions become statistically time-independent on the convective timescale.  The approach to dynamical equilibrium is shown in \Fig{Ut-vs-time}.  In that figure, we plot the magnitude of the horizontal velocity component $u_\mathrm{h}$ half a pressure scale height ($H_p = -pdr/dp$) below and above the peak in the Brunt-V\"ais\"al\"a (BV) frequency squared, $N^2$ (\Eq{BV}), that marks the convective boundary. Although there is noise, it is clear from this figure that dynamical equilibrium is established after a time of about 400 h. While in dynamic equilibrium, the mass entrainment rate slowly decreases as the simulation approaches a state closer to thermal equilibrium. The entrainment analysis can be found in \Sect{entrainment} using the same methodology as in \paperone.

The radius of the CB is marked by a fairly sharp peak in the Brunt-V\"ais\"al\"a (BV) frequency. In the equations below, we decompose $N^2$, the square of the BV frequency, into contributions $N_t^2$ and $N_\mu^2$ that arise from the temperature and compositional gradients, respectively. A positive $N^2$ indicates stability, suppressing convective processes, and a negative $N^2$ implies instability to convection.  A sharp peak in $N^2$ at the CB therefore strongly impedes any residual convective motions there (see \Fig{Prad-1000xht-Krange}).

\begin{eqnarray}
        N^2 &=& \frac{g \delta}{H_p}(\nabla_\mathrm{ad}-\nabla_{\rm star})+\frac{g \delta}{H_p}\frac{\phi}{\delta}\nabla_\mu \lEq{BV}  \\
        N_t^2 &=& \frac{g \delta}{H_p}(\nabla_\mathrm{ad}-\nabla_{\rm star}) \lEq{BV_t} \\
        N_{\mu}^2 &=& \frac{g \delta}{H_p}\frac{\phi}{\delta}\nabla_\mu  \lEq{BV_mu}
\end{eqnarray}

where

\begin{eqnarray}
       & \delta = -(\frac{\partial ln \rho}{\partial ln T})_{p,\mu} , \phi = (\frac{\partial ln \rho}{\partial ln \mu})_{p, T}  \nonumber \\
       & \nabla_{\rm star} = \frac{dln T}{dln p}, \nabla_\mathrm{ad} = (\frac{dln T}{dln p})_S, \nabla_\mu = \frac{dln \mu}{dln p} \nonumber
\end{eqnarray}

Here $\rho$ is density, $T$ temperature, $H_p$ pressure scale height, $\mu$ mean molecular weight, $S$ specific entropy, $\nabla_{\rm star}$ the actual temperature gradient, $\nabla_\mathrm{ad}$ the adiabatic gradient, and $\nabla_{\rm star}-\nabla_\mathrm{ad}$ is the superadiabaticity.

In ste7llar  evolution models, the CB is usually defined as the radius at which the adiabatic gradient is equal to the radiative gradient, also known as the SB. Based on our discussion of a very long-duration simulation in \Sect{long-term}, we choose to define the CB in this work as the radius where, in statistical dynamical and thermal equilibrium, the radial derivatives of the radiative and convective heat fluxes as well as the convective heat flux itself and the kinetic energy dissipation rate all vanish. The CB, thus defined, is different from the SB, because at the SB the radial derivative of the radiative heat flux does not vanish.

\begin{figure}
  \includegraphics[width=\columnwidth]{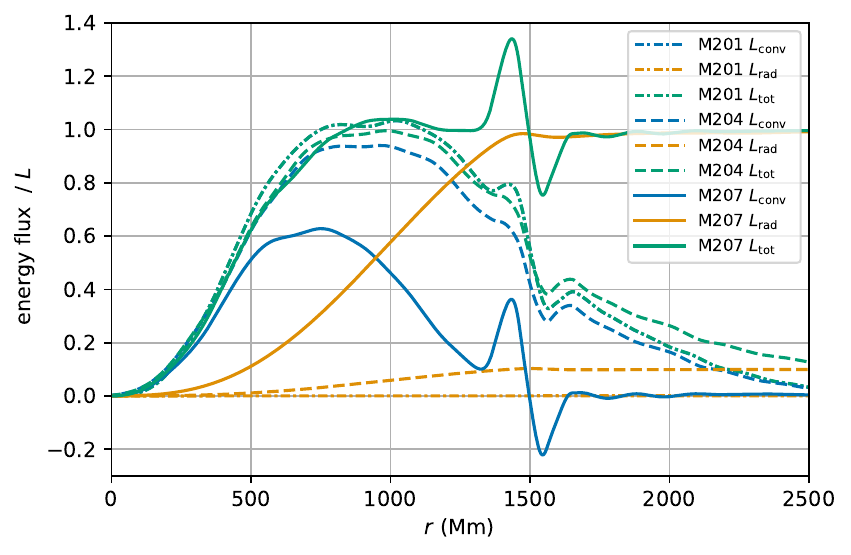}
  \caption{
Total convective (radiative) energy flux normalized by the luminosity for the no-diffision (M201, 0x), intermediate-diffusion (M204, 100x) and high diffusion (M207, 1000x) simulations with 1000x luminosity enhancement at the same data dump (2556.6, 2470.7 and 2470.7 h), all with grids of $1152^3$ cells. The curves are smoothed by using moving averages three times over a window 120 \Mm\ wide and time-averaged over 100 dumps $\sim\ 70\ \mathrm{hr}$.
The fluxes are defined in \Eq{conv-flux} and \Eq{rad-flux}. Temperature, opacity, and density in \Eq{rad-flux} are spherical averages.}
    \lFig{flux}
% This figure is produced by gradM200.ipynb.
\end{figure}

\begin{figure}
       \includegraphics[width=\columnwidth]{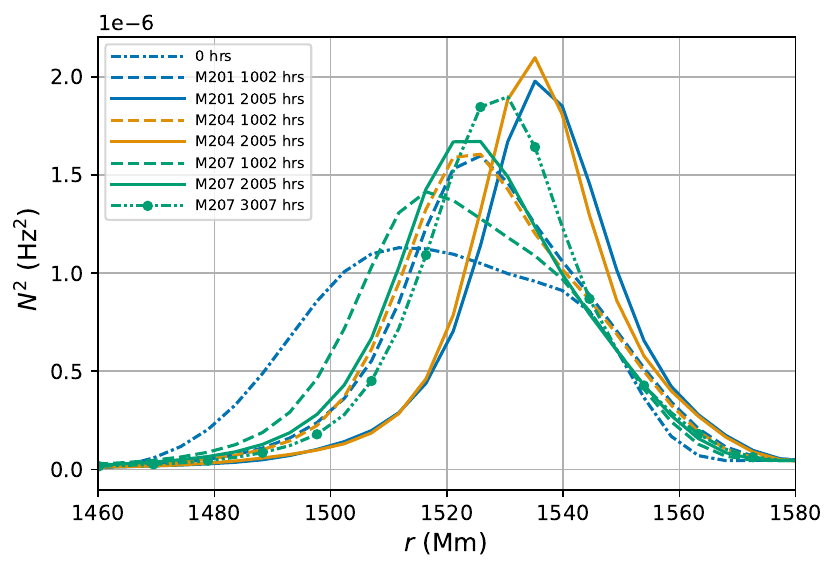}
       \includegraphics[width=\columnwidth]{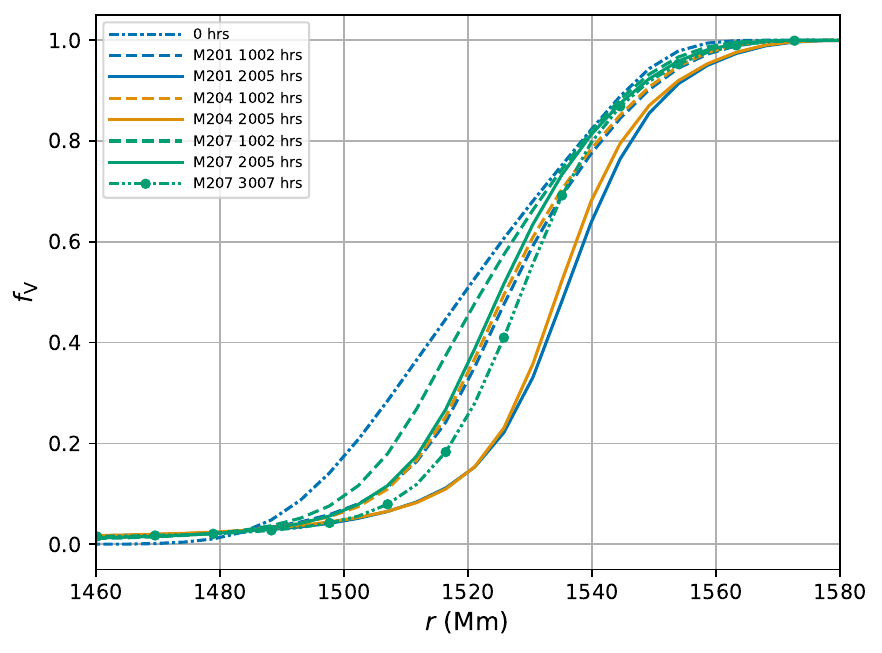}
       \includegraphics[width=\columnwidth]{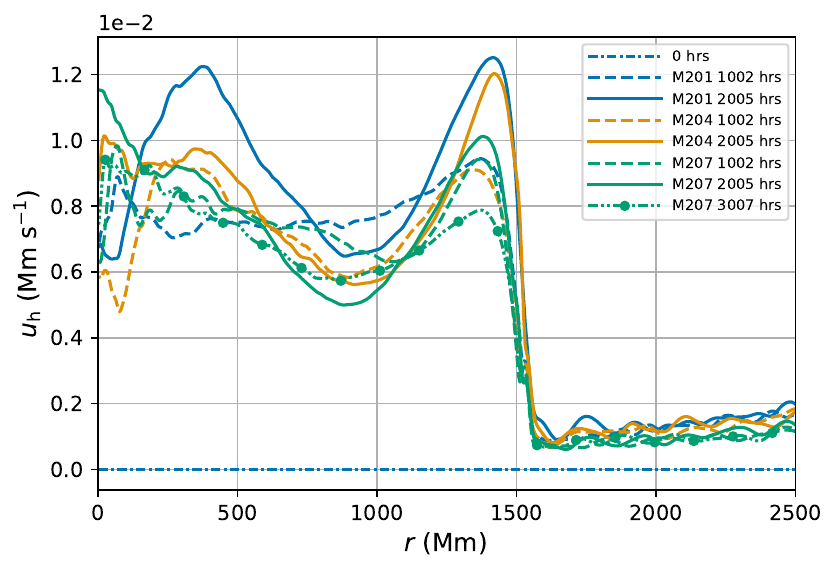}
       \caption{Profiles of $N^2$, $f_{\rm V}$, the horizontal velocity magnitude $u_\mathrm{h}$ of the simulations with 1000x luminosity enhancement (M201: no diffusion, M204: $k \sim L^{2/3}$, M207: $k \sim L$, all at the $1152^3$ grid resolution.)}
    \lFig{Prad-1000xht-Krange}
% This figure is produced by gradM200.ipynb.
\end{figure}

\subsection{Dynamics and kinematics in dynamical equilibrium}
The convection rapidly organizes itself such that the total convective flux becomes the luminosity minus the total radiative energy flux (\Fig{flux}, \Eq{conv-flux}, \Eq{rad-flux}). Therefore, our simulated star reaches a dynamical equilibrium over the first few convective turn-overs and stays in dynamical equilibrium thereafter.

\subsubsection{Effect of radiative diffusion}
\Fig{Prad-1000xht-Krange} shows how $N^2$, $f_{\rm V}$ and $u_\mathrm{h}$, evolve for different strengths of radiative diffusion at 1000x the nominal luminosity.
 Outward from the SB by about 120 \Mm\ ($10\%$ in radius) in the initial state of the simulation, $N^2$ has a strong, slowly migrating peak reflecting the sudden change of entropy mainly caused by the change in $\mu$ at that location.

Perhaps the most important effect of the radiative diffusion is that, as this is increased, the position of the composition change, traced by the $f_{\rm V}$ profile, moves outward less rapidly. This effect can also be seen in the position of the $N^2$ peak feature. This behavior can be explained by the fact that when we add radiative diffusion, we introduce into the problem a mechanism for carrying the heat introduced into the convection zone outward through the stably stratified envelope.  In the absence of this mechanism, in addition to entraining high entropy materials from the envelope, heat must pile up in the convection zone, and this must cause it to expand. This is analogous to the helium shell flash in that the ignition of helium fusion in a thermal pulse produces more energy temporarily than can be carried away by radiative diffusion, causing the star to expand and brighten \citep{Herwig:2006gk}.
In our high diffusion case, heating by nuclear burning is, on average, removed by the heat energy flowing through the reflecting sphere at our outer boundary in the form of radiation (\Fig{flux}). The total convective flux and total radiative energy flux are calculated by \Eq{conv-flux} and \Eq{rad-flux}.

\begin{eqnarray}
       &L_{\mathrm{conv}}(r)=\iint\limits_{\mathrm{sphere}\ r} (p +\rho e+ \frac{1}{2}\rho u^2) u_r dA  \lEq{conv-flux} \\
       &L_{\mathrm{rad}}(r)=-\frac{4 \pi r^2 c}{3\kappa\rho}\frac{\partial (a T^4)}{\partial r}  \lEq{rad-flux}
\end{eqnarray}

The convective flux is the flux of enthalpy plus the kinetic energy summed over the sphere at radius $r$. $c$ is the speed of light.

 In cases of no diffusion, there is no diffusive heat flux across the stably stratified gas in the outer part of our computational region. The heated convective core pushes the envelope resulting in positive convective flux at all radii. We measure that about $55\%$ of the nuclear heating becomes potential energy by expanding the convective core and compressing the stable envelope (i.e. redistributing mass in a static gravitational potential), while $45\%$ becomes internal energy by heating the star up. In the intermediate diffusion case M204, $42\%$ of the nuclear heating expands the core and $46\%$ heats the star up. About $10\%$ of the nuclear heating is transported outward by radiative diffusion in that case.
In \Fig{Prad-1000xht-Krange}, the convective velocity is slightly smaller in the high diffusion case but the profile of the magnitude of horizontal velocity remains similar. In all cases, the kinetic energy is negligible. Once a dynamical equilibrium is established, it mostly does not change over time and stays negligible. The effect on the motion in the stable envelope, i.e., IGWs, is discussed in \Sect{quasi-steady}.

The differences in the heights and shapes of the $N^2$ peaks, between the cases of no diffusion and intermediate diffusion at the same time (1000 or 2000 hours), are very small (\Fig{Prad-1000xht-Krange}), because most of the heat injected (90\% and 100\%) piles up in the convective core, which leads to quantitatively similar dynamics. However, in the case of high diffusion, the change of location and shape in the $N^2$ peak is noticeably smaller than in the other two cases given the same amount of time (\Fig{Prad-1000xht-Krange}). However, the overshoot and undershoot of the convective flux, and the overshoot of radiative flux at 1500 \Mm\ suggest the thermal structure is adjusting, at a small rate.
Hence, any significant change in the stratification for the high diffusion cases happens on a longer timescale than no or intermediate diffusion.
To reduce the computational cost of studying the evolution on a longer timescale, we investigate the effect of enhancing luminosity in the next section and the possibility of accelerating the evolution by enhancing the luminosity in \Sect{long-term}.

The heat piling up in the no or intermediate diffusion cases explains the fact that the star lifts the convective core and compresses the envelope. This process will continue and completely change the stratification because the total energy of our simulation keeps increasing in these two cases. Hence, to simulate a realistic star in thermal equilibrium, the only reasonable scenario is the high diffusion one, and we later discuss the effect of enhancing luminosity using the high diffusion cases only. In addition, as discussed in \Sect{entrainment}, the entrainment continues at a relatively constant rate, which suggests that the star is still adjusting its stratification and has not yet reached a thermal equilibrium. In such an equilibrium, all the temporal dependence on time scales longer than several large eddy turn-overs in the convection zone could be expected to very nearly vanish. By definition, the total heat content will be radiated away at the rate of the luminosity on a thermal timescale, if there is no nuclear heating. Therefore, it is not feasible to investigate the dynamics on a thermal timescale in the cases of no or intermediate diffusion without disrupting the thermodynamical structure completely. Hence, the discussion on the evolution on a thermal timescale in \Sect{long-term} and \Sect{penetration} focusses on the high diffusion cases.

\subsubsection{Effect of enhancing luminosity}
\begin{figure}
       \includegraphics[width=\columnwidth]{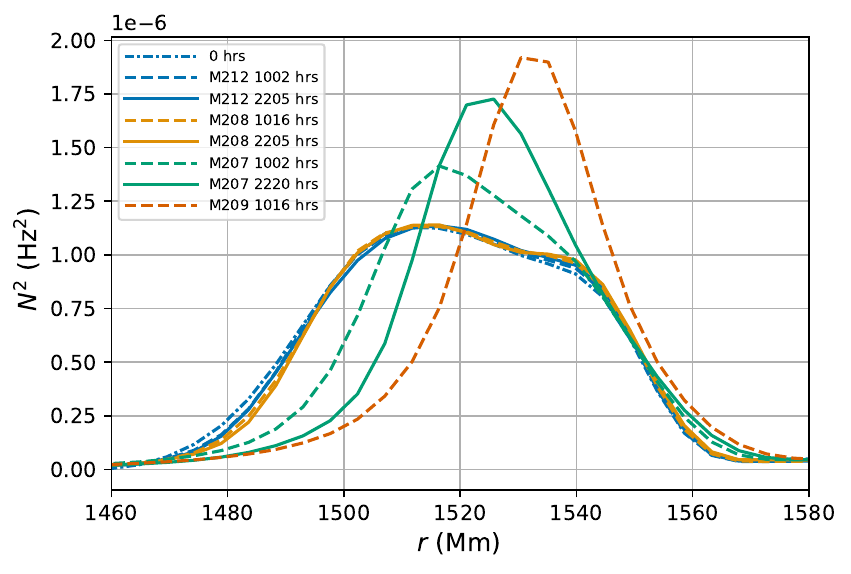}
       \includegraphics[width=\columnwidth]{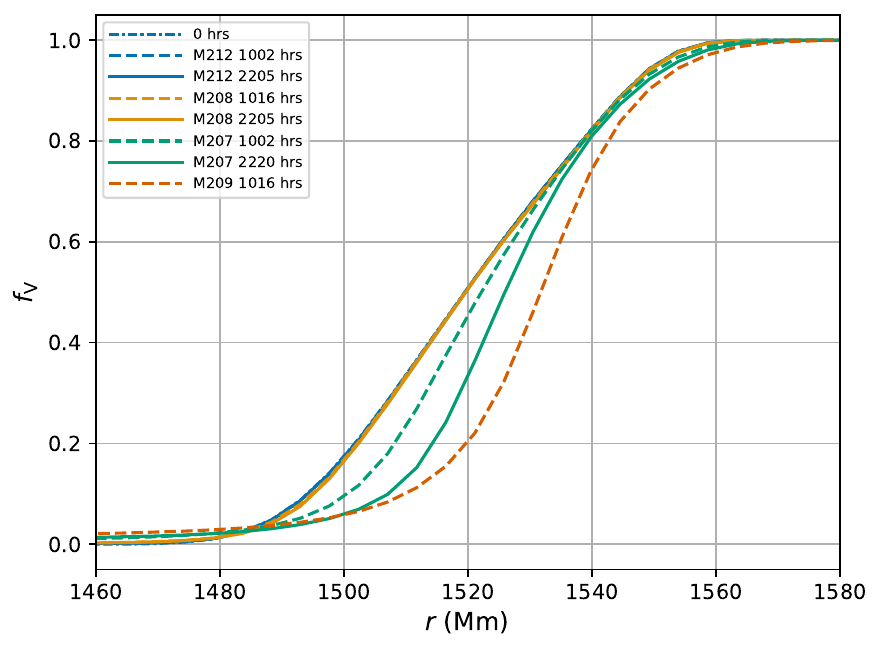}
       \includegraphics[width=\columnwidth]{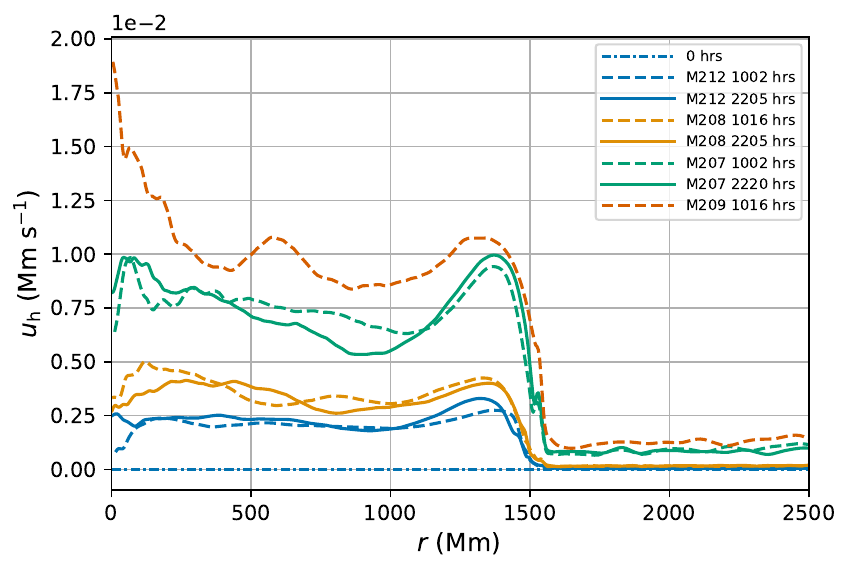}
       \caption{Profiles of $N^2$, $f_{\rm V}$, $u_\mathrm{h}$ of M212 (31.62x $L_*$ \& $k_*$), M208 (100x $L_*$ \& $k_*$), M207 (1000x $L_*$ \& $k_*$), M209 (3162x $L_*$ \& $k_*$), all at the $1152^3$ grid resolution.}
    \lFig{Prad-K-L}
% This figure is produced by gradM200.ipynb.
\end{figure}

\Fig{Prad-K-L} shows the profiles of $N^2$, $f_{\rm V}$, and horizontal velocity for a series of runs in which we vary the luminosity. For each boosting factor, we also enhance the radiative diffusion by the same factor.  Cases of enhancement factors of 31.62, 100, 1000, and 3162 are used. For the two lowest luminosity cases, we observed essentially no change within 2000 hours in the profile of $f_{\rm V}$, and in the position and the shape of the $N^2$ peak during these simulations.
This certainly does not mean that changes would not result were these two simulations run longer in time.

Runs M207 and M209, with luminosity enhancement factors of 1000 and 3162, reshape the initial $f_{\rm V}$ radial profile within relatively short times of less than 2500 hours. After this intial reshaping in these high-power cases, the $f_{\rm V}$ radial profile translates while maintaining its shape as the gas from above the convection zone is entrained. As will be discussed in \Sect{long-term}, boosting the nuclear heating and the radiative diffusion by a common factor can be regarded as accelerating the time rate of change of the stellar model. In order to probe the long-time behavior of the stellar model, this balanced enhancement of the luminosity and radiation diffusion is appealing for our explicit \code{PPMstar} code, because it dramatically lowers the cost of finding the long-time behavior.

\begin{figure}
       \includegraphics[width=\columnwidth]{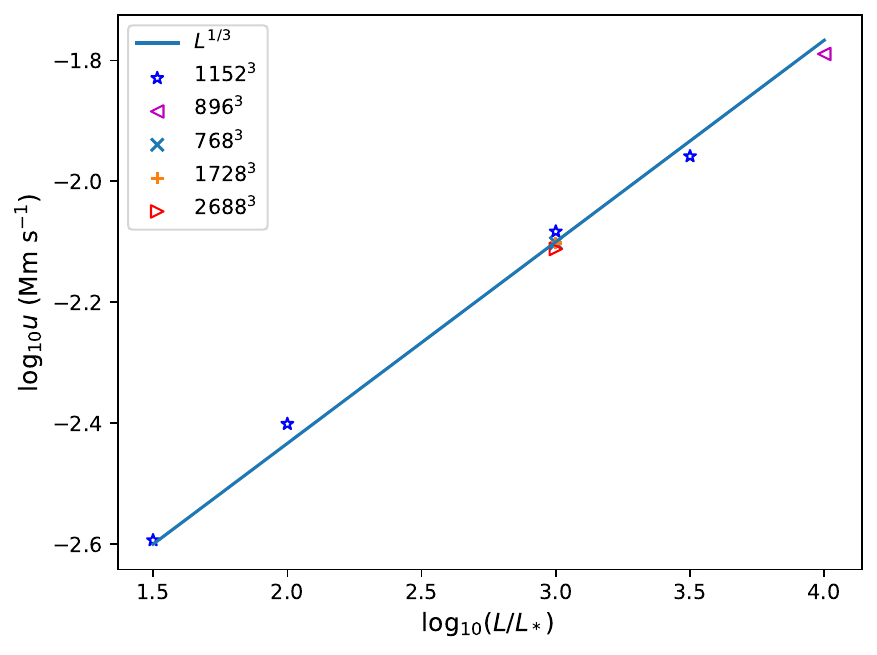}
       \caption{Luminosity versus convective velocity magnitude in the convection zone at 1000 \Mm\ averaged over 20 data dumps. All cases are high diffusion.}
    \lFig{u-L}
% This figure is produced by gradM200.ipynb.
\end{figure}
\Fig{u-L} confirms that the magnitude of velocity scales with $L^{1/3}$ in the presence of radiation pressure and radiative diffusion. This scaling is also observed in \paperone.

\subsubsection{Convergence}
\begin{figure}
       \includegraphics[width=\columnwidth]{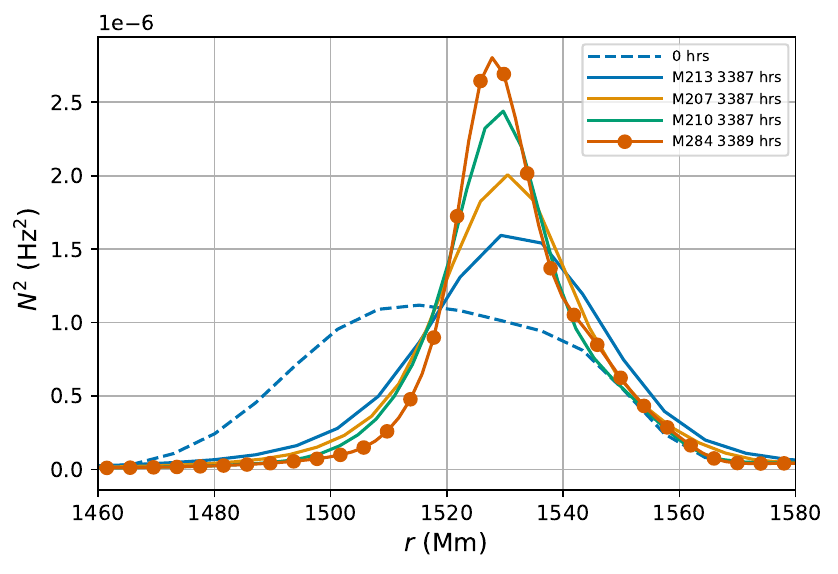}
       \includegraphics[width=\columnwidth]{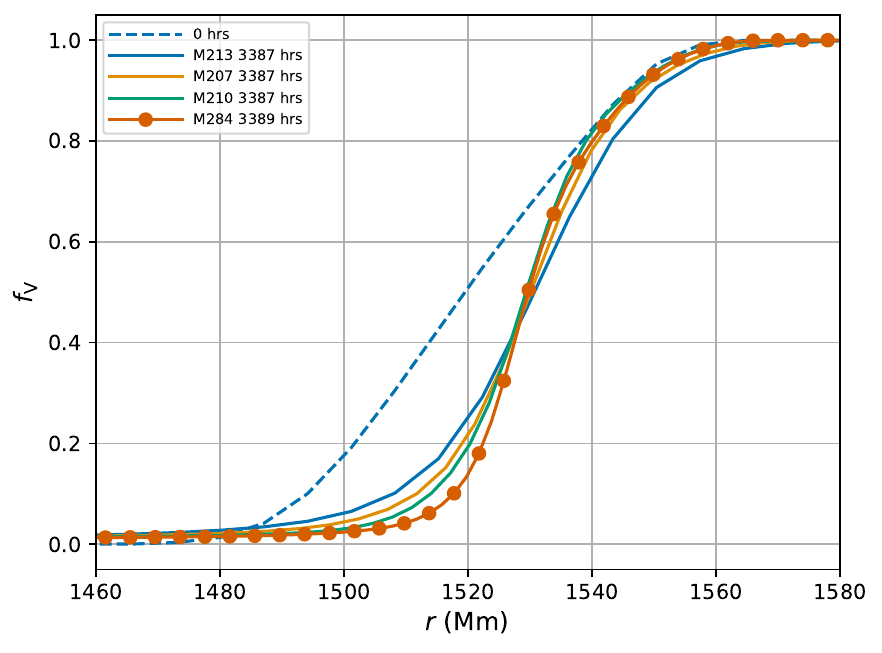}
       \includegraphics[width=\columnwidth]{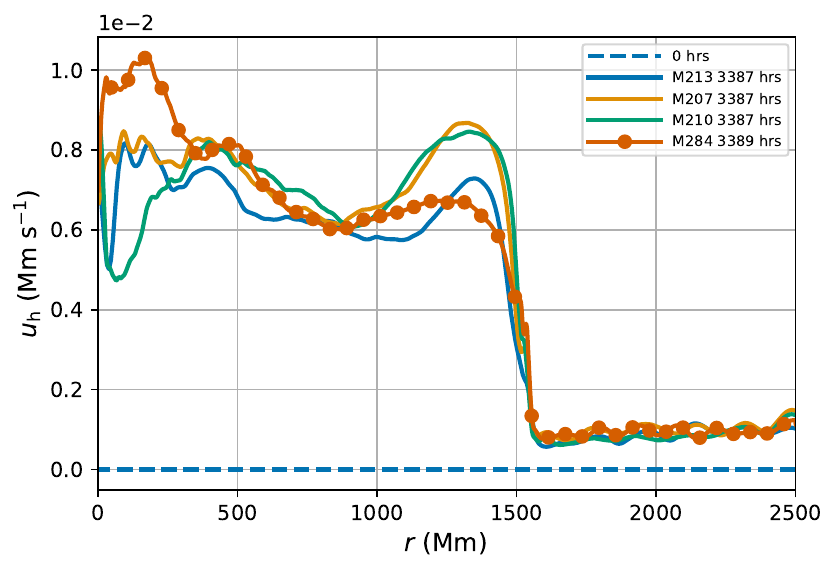}
       \caption{Snapshots of $N^2$, $f_{\rm V}$, $u_\mathrm{h}$ of M213: $768^3$, M207: $1152^3$, M210: $1728^3$, M284: $2688^3$, all with 1000x $L_*$ and 1000x $k_*$.}
    \lFig{Prad-1000x-K-L-res}
% This figure is produced by gradM200.ipynb.
\end{figure}
In \Fig{Prad-1000x-K-L-res}, the profiles of $N^2$, $f_{\rm V}$ and horizontal velocity are presented for a sequence of simulations carried out at different grid resolutions to show the effect of refining our computational grid. These simulations are performed with a luminosity and radiation diffusion enhancement factor of 1000. We see that the $N^2$ peak becomes higher with increasing grid resolution. However, the location of the $N^2$ peak is roughly the same regardless of the resolution.  The radial profile of $f_{\rm V}$ becomes steeper with grid refinement, and it is clear that this steepening is not complete even on the highest resolution grid shown in the figure.
Although there is some statistical noise evident in the plots of the horizontal component of the velocity in \Fig{Prad-1000x-K-L-res}, it is clear that these simulations have converged upon mesh refinement to a well-defined state.  Even the radial profiles of $f_{\rm V}$ near the CB appear to have converged in terms of the position of the sharp increase in $f_{\rm V}$ though not in its steepness. The interpretation of the $N^2$ peak and the slope of the $f_{\rm V}$ not converging on grid refinement is that we have not converged on mixing. In \Sect{long-term}, convergence will be shown for turbulent dissipation measured from the simulations and for vorticity in the stable envelope.

\subsubsection{Mixing length parameter}
\lSect{MLT}
We first check the efficiency of convection. The mean free path of a photon inside our star is of order of $1-10\ \cm$, i.e.\ our star is opaque and radiative transport of energy can be treated as a diffusion process. We take $r_c=1500\ \Mm$ as the radius of our convective core, the thermal adjustment timescale of the convective core will be $\tau_\mathrm{t} = r_{c} ^2\rho c_p/k$. The convective timescale is
$\tau_{c}=2 r_{c}/v_{c}\ .$ For our M207 case, the boosting factor for the radiative diffusion can be interpreted as increasing
the thermal conductivity by a factor of 1000. Given that, 
\[
  \frac{\tau_{c}}{\tau_\mathrm{t}}=\frac{2 k}{v_{c}r_{c}\rho c_p} \approx \natlog{1.06}{-4}
\]
 the convection in our simulations is efficient in transporting excess heat.
We measure the super-adiabatic temperature gradient in our simulations and hence can derive numerical values of the standard mixing length parameter $\alpha$.

In MLT, the total convective flux is modeled as linearly proportional to $\alpha^2$  \citep{prialnik2000introduction}, the square of the mixing length parameter:
\begin{equation}
     L_{\mathrm{conv}} = 4\pi r^2 \rho c_p \sqrt{p/\rho}(\nabla_{\rm star}-\nabla_{\mathrm{ad}})^{3/2} \alpha^2
    \lEq{mlt_flux} 
\end{equation}
where symbols have their usual meanings.

From the superadiabaticity in \Fig{supad}, we see that the temperature gradient is nearly adiabatic throughout the convective core ($\nabla_{\rm star} -\nabla_\mathrm{ad} \sim\ 10^{-4}$). The convective core is slightly superadiabatic inside 1000 \Mm\ and becomes slightly subadiabatic beyond 1000 \Mm. This is where the radial entropy gradient $dS/dr$ becomes positive and the convective flows start to encounter the marginally stable stratification. Though the convective stability criterion indicates the stratification is stable at 1000 \Mm\ and outward, this slightly subadiabatic temperature gradient cannot bring the convetive motion to a halt. The flows continue before arriving at the very much more significant entropy gradient at the CB.

The convective flux is propotional to $(\nabla_{\rm star}-\nabla_\mathrm{ad})^{3/2}$. However, the temperature gradient is not superadiabatic throughout the entire convective core (\Fig{supad}). If we take the approach in \cite{Porter:2000kv}, redefining the superadiabaticity as $\nabla_{\rm star} - f\nabla_{\mathrm{ad}}$
where $f=0.999$, we find that the entire convective core is superadiabatic and the mixing length parameter $\alpha$, solved from \Eq{mlt_flux}, is in the range from 0.4 to 1.2 (\Fig{asq}). This value of $f$ is different from the value
0.98 used in \cite{Porter:2000kv}.  \cite{chan1989turbulent} suggest that the superadiabaticity might depend on the aspect ratio of the convective spherical shell and upon the equation of state.
We find that $\nabla_{\rm star} - \nabla_\mathrm{ad}$ is positive inside 1000 \Mm\ and negative beyond 1000 \Mm\ for all our different heating rates, but its magnitude increases with the boosting factor. This is qualitatively in agreement with the MLT assertion that the convective flux scales with superadiabaticity to the power of $3/2$.

\begin{figure}
  \includegraphics[width=\columnwidth]{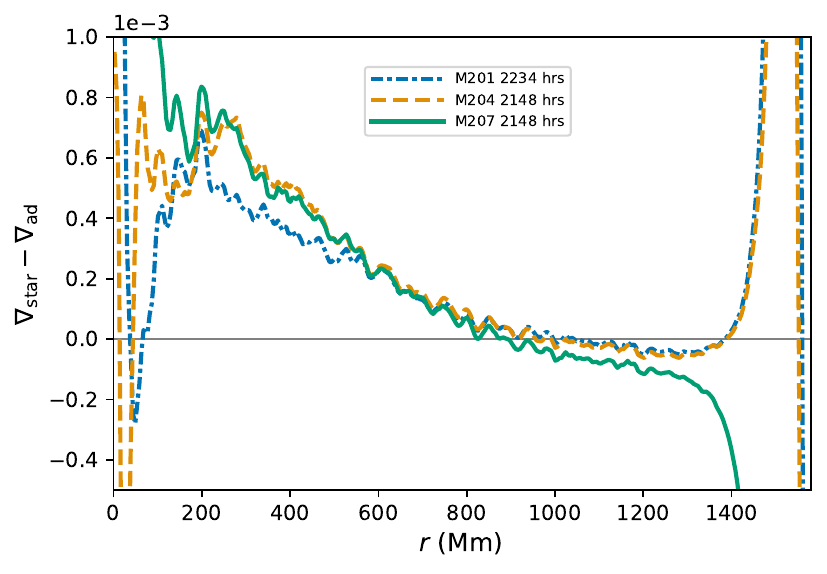}
  \caption{The superadiabaticity ($\nabla_{\rm star} - \nabla_\mathrm{ad}$) of M201 (1000x heating, 0x diffusion), M204 (1000x heating, 100x diffusion) and M207 (1000x heating, 1000x diffusion), averaged over 100 dumps ($\sim$ 70 hours) and 30 \Mm.}
  \lFig{supad}
% This figure is produced by gradM200.ipynb.
\end{figure}

\begin{figure}
  \includegraphics[width=\columnwidth]{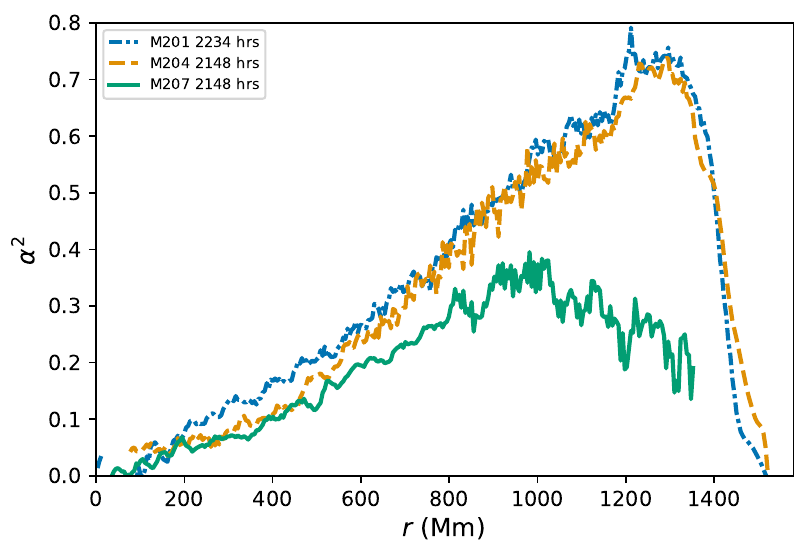}
  \caption{The mixing length parameter squared $\alpha^2$ of M201, M204, M207 averaged over 100 dumps ($\sim$ 70 hours) and 30 \Mm.}
  \lFig{asq}
% This figure is produced by gradM200.ipynb.
\end{figure}

\subsection{Mass entrainment rate}
\lSect{entrainment}
\begin{figure}
  \includegraphics[width=\columnwidth]{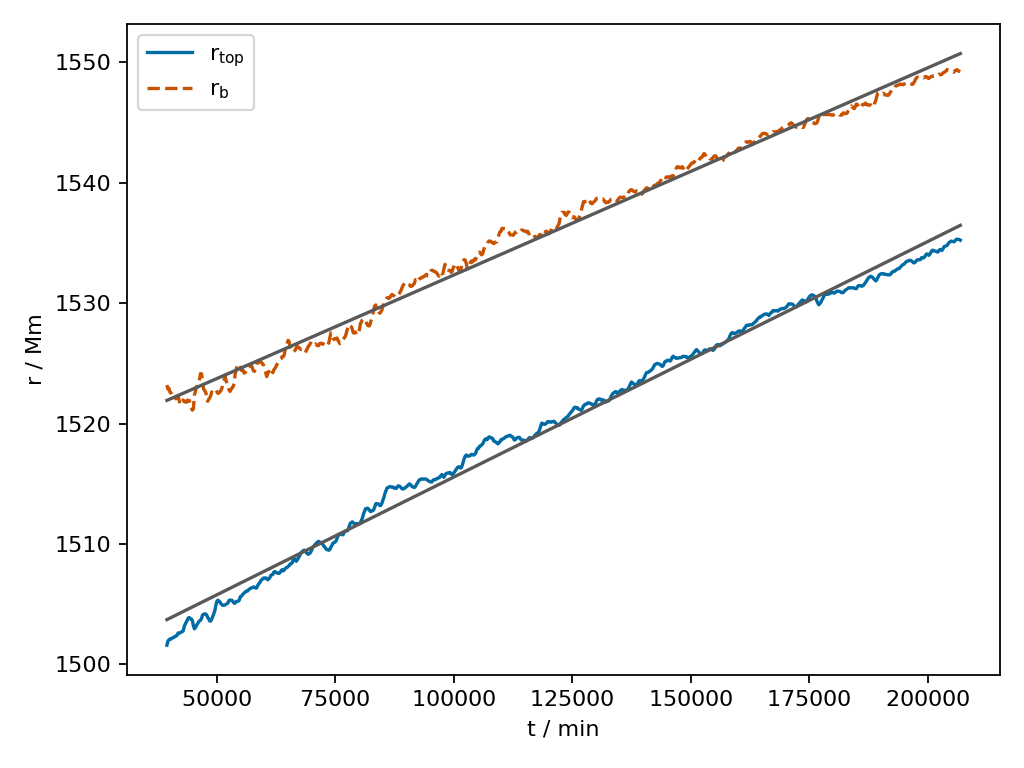}
  \includegraphics[width=\columnwidth]{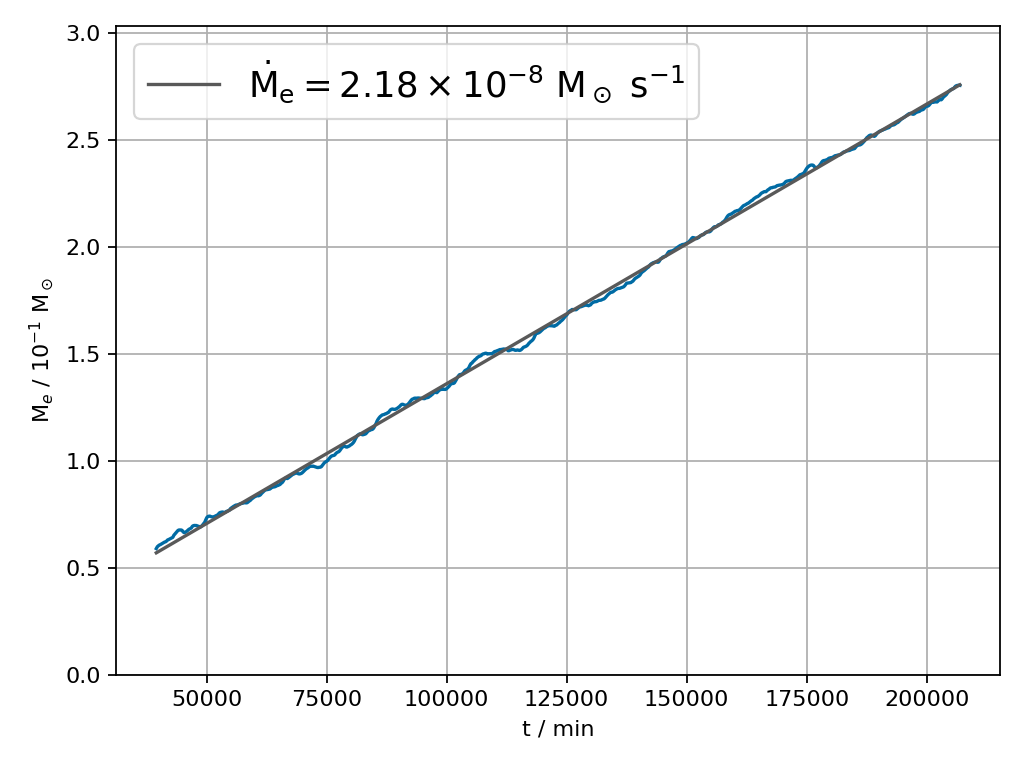}
  \caption{The time evolution of the radius of maximal $df_{\rm V}/dr$ minus one $f_{\rm V}$ scale height, and the entrained masses of M201.}
  \lFig{ent}
\end{figure}
We determine the entrainment rate of the envelope gas from above the CB into the convection zone using the same methodology as in \paperone. As in \paperone\  we define the entrained mass as the total mass of the envelope material within $r_{b}$. $r_{b}$ is the location of the maximum gradient of $f_{\rm V}$ less one $f_{\rm V}$ scale height. This entrained mass evolves linearly with time, and one example is shown in \Fig{ent}.
% \textcolor{Green}{(To Falk: I don't know how I can re-configure the entrainment and % CB plots from local\_methods.entrainment.analysis.)}

% The entrainment rates of M114 ($P_{gas}$, no diffusion, 1000x heating) and M201 ($P_{gas}+P_{rad}$, no diffusion, 1000x heating) are measured from about the same period of time. By including radiation pressure, the entrainment rate reduces to \unit{\natlog{6.85}{-1}}{\Msun/\yr} from \unit{\natlog{7.98}{-1}}{\Msun/\yr}.
Compared to the  $P_\mathrm{gas}$ only case (M114 in \paperone) the entrainment rate is $14\%$ smaller when adding $P_{\mathrm{rad}}$ (M201), and decreases by $50\%$ when also adding radiative diffusion (M207).
% \unit{\natlog{2.17}{-12}}{\Msun/\second} from \unit{\natlog{2.53}{-12}}{\Msun/\second}.

We estimate the entrainment rate at nominal heating by extrapolating separately from three sets (no, intermediate and high diffusion) of simulations (\Fig{entrainment}). The entrainment rates for no diffusion and intermediate diffusion are practically the same. The difference between the entrainment rates extrapolated from these two sets are due to the uncertainty of the fitting slope.

The extrapolated entrainment rate cannot persist for a significant fraction of the main-sequence lifetime  (\Sect{long-term}). We believe instead that the large entrainment rates that we observe after our simulations initially establish a dynamical equilibrium, are the result of thermal non-equilibrium. We will discuss the development of penetrative convection on a longer time scale and the effect on the entrainment of the resulting subadiabatic temperature gradients within the penetrative region between the SB and the CB in \Sect{long-term} and \Sect{penetration}
\begin{figure}
       \includegraphics[width=\columnwidth]{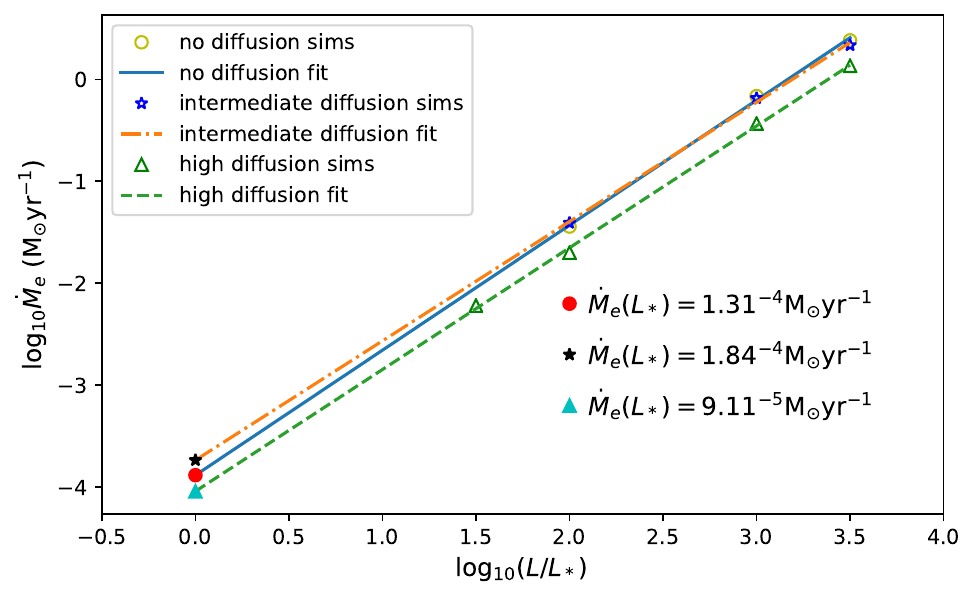}
       \caption{Entrainment rates of simulations (hollow symbols); extrapolated entrainment rates at nominal
       heating (solid symbols).}
    \lFig{entrainment}
% from EntrainmentExtrapolation.ipynb
\end{figure}

\subsection{IGWs}
\lSect{IGWs}
One important consequence of radiative diffusion is damping of IGWs in the stably stratified layers of the star \citep{Zahn:97}.  We study this effect of radiative diffusion in our model star by observing the wave motions in the envelope surrounding the convective core. Using the same approach as in \papertwo, we decompose the radial component of the velocity field into complex spherical harmonics coefficients using the \code{SHTools} package \citep{Wieczorek:2018}. We then perform a Fourier transform on each coefficient time-series. Then we use the \code{SHTools} package to calculate the power spectral density of the radial velocity oscillations normalized by degree $\ell$ for each frequency bin. The time interval between data dumps in our simulations determines an upper limit to the frequencies that we can observe.  This upper limit is about 200 ${\rm \mu}$ \Hz\ for the simulations reported here, corresponding to $\approx 43\ \minute$ between dumps. In these simulations, we have located our outer boundary so that the radius of the convective core is about $60\%$ of the boundary radius. The degree $l$ of the spherical harmonics gives the number of nodes going along a meridian from one pole to the other. Hence at the CB ($60\%$ of the maximal radius in our computational region), with a $1152^3$ grid, we can resolve, in principle, spherical harmonics up to $l=\frac{\pi r_{\rm CB}}{4\cdot \Delta x} \approx\ 250$, where $r_{\rm CB}$ is the radius of the CB and $\Delta x$ the cell width, because the data we use in this analysis has been averaged over cubical bricks of grid cells 4 on each side before being written to disk \citep{stephens:21}.

As shown by the velocity profile in \Fig{Prad-1000xht-Krange}, the convection in the core is less vigorous (smaller $u$) 
in high diffusion. Therefore, the excitation of IGWs \citep{Edelmann:2019jh} becomes less efficient due to radiative diffusion. Radiative diffusion damps both the IGWs and the excitation of IGWs, resulting in the power spectra we observe.

As shown in \Fig{1000xht-k-w} for the radial velocity component, most of the power of the wave motions  is concentrated at frequencies below the maximum
Brunt-V\"ais\"al\"a (BV) frequency in the stable envelope (see also \papertwo). It is also concentrated in $l's$ smaller than 80.
Modes with small-scale structures $l>100$ are damped in simulations with high diffusion, and less so in intermediate diffusion. \Fig{0kvs100k-damping} shows the damping effect in terms of power ratio of M204 and M207 to M201. Modes of $l>80$ are reduced in power by more than $95\%$ in high diffusion and by $50\%$ in intermediate diffusion. However, for the more important frequencies below the BV frequencies, radiative damping in high-diffusion simulations reduces the wave amplitudes by a factor 2.5 to 5.

In \paperone, a formula is considered for predicting the diffusion coefficient that might produce material mixing in the stably stratified envelope due to IGW-induced motions.  According to that relation the diffusion coefficient should scale with the square of the vorticity in the envelope, among other factors.  In that study, working with simulations without radiative diffusion, it was found that this envelope vorticity shows no sign of convergence under grid refinement. The power spectra in \Fig{0kvs100k-damping} show that radiative damping of the high $l$ IGW modes in our high diffusion cases allows the vorticity in the envelopes of these simulations to converge with mesh refinement. In \Fig{vortRes}, the vorticity in the envelope does not change when the grid is refined in the presence of radiative diffusion.

\begin{figure}
  \includegraphics[width=\columnwidth]{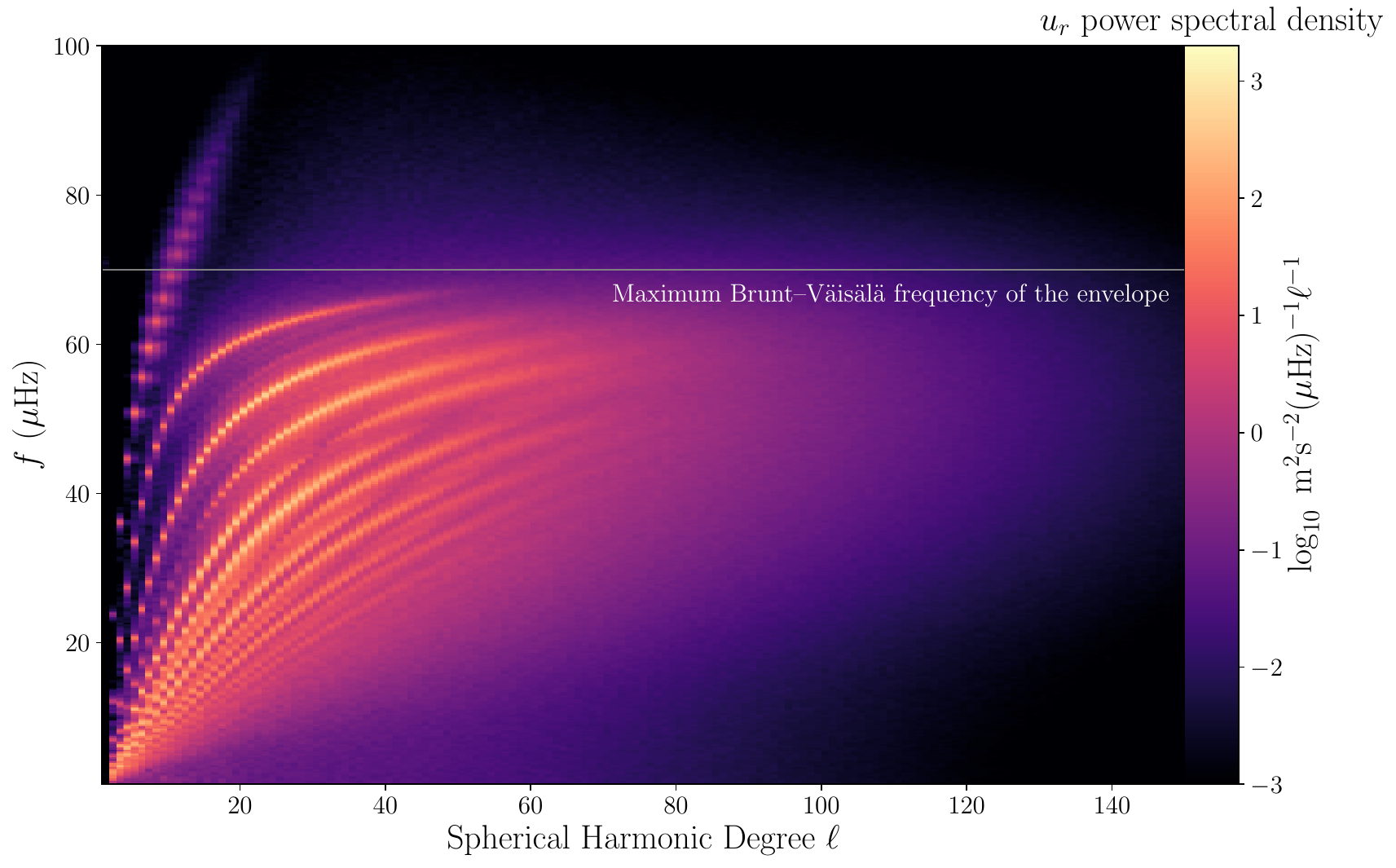}
  \includegraphics[width=\columnwidth]{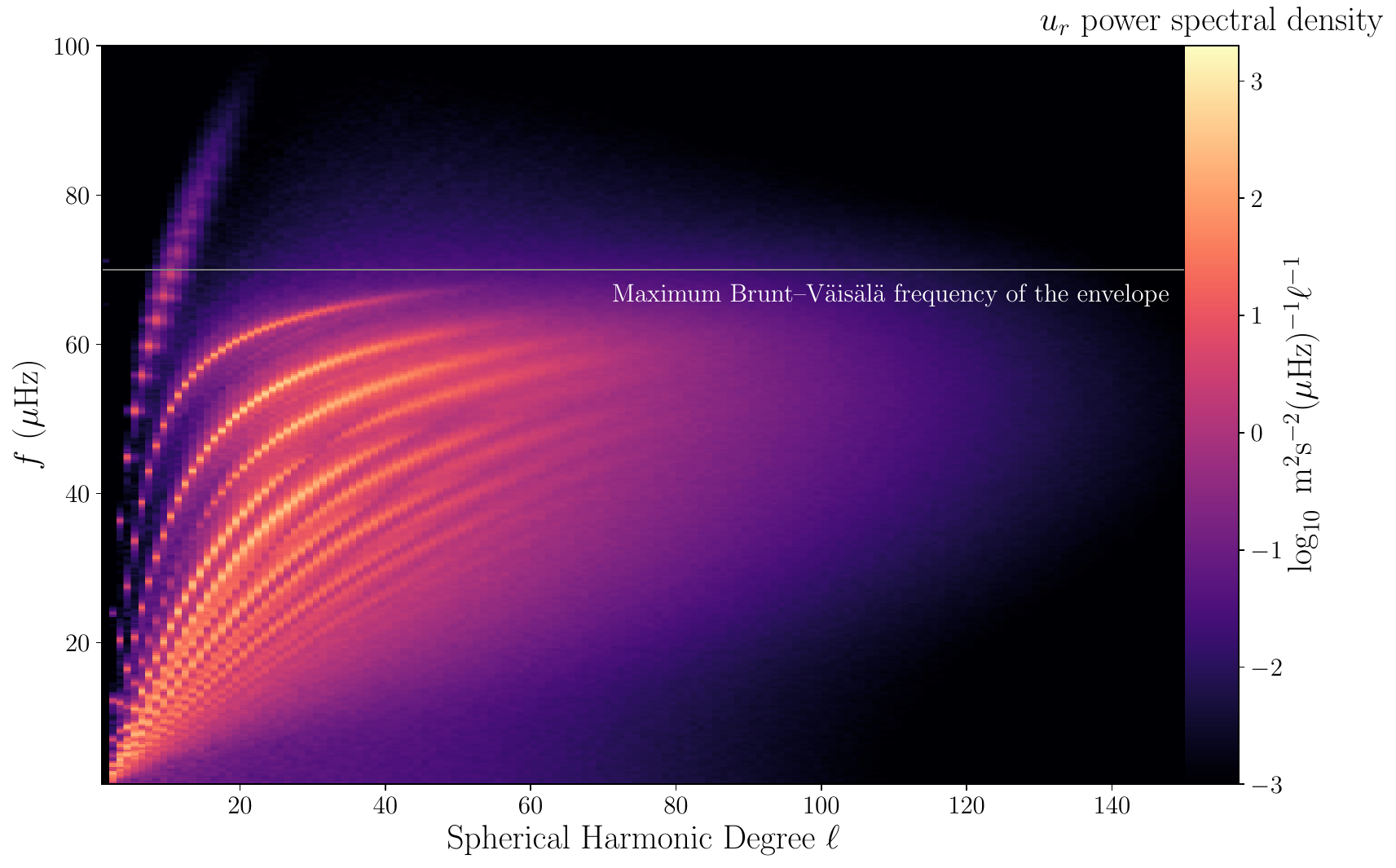}
  \includegraphics[width=\columnwidth]{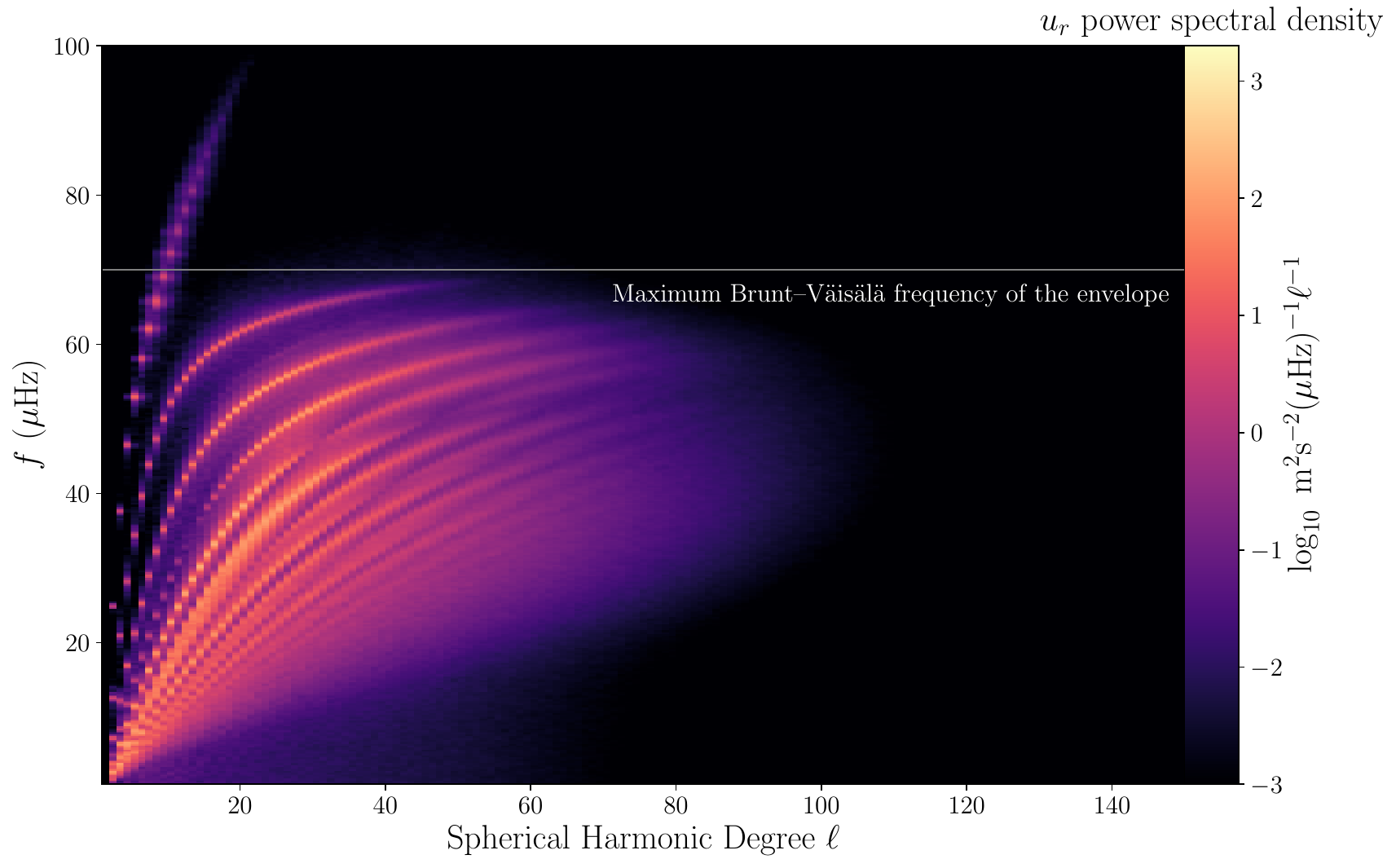}
  \caption{Spectral power density of $u_r$ at $19\Msun$ of M201 (top, 1000x heating, 0x diffusion), M204 (center, 1000x heating, 100x diffusion) and M207 (bottom, 1000x heating, 1000x diffusion) averaged over $\sim$\ 1160 hr centered at 2556.6 hr).
}
%Notebook for k-omega without axis spectra: http://206-12-89-164.cloud.computecanada.ca/sefffal/hydro-asteroseismology/-/blob/master/notebooks/komega_ur_M117.ipynb
  \lFig{1000xht-k-w}
\end{figure}

The amount of radiative damping of the IGWs in the envelope is of interest when we consider the possibility that these IGWs cause material mixing in the envelope.  The short wavelength waves that are damped substantially, as seen in \Fig{1000xht-k-w}, have no effect upon the asteroseismology observations of the waves at the stellar surface of massive stars, as they would be located in the region of white noise \citep{Bowman2020b}. However, it is possible that the short wavelength waves have a significant impact on the efficiency of material mixing. This potential for IGW envelope mixing is explored at length in \paperone.  Here we see that the short wavelength waves are damped by radiation diffusion. It is generally believed that radiative diffusion can play an essential role in IGW-induced mixing (e.g. \cite{Townsend:1958}, \cite{zahn:74}, \cite{press:81}, \cite{garaud:17}, \paperone).

\begin{figure}
  \includegraphics[width=\columnwidth]{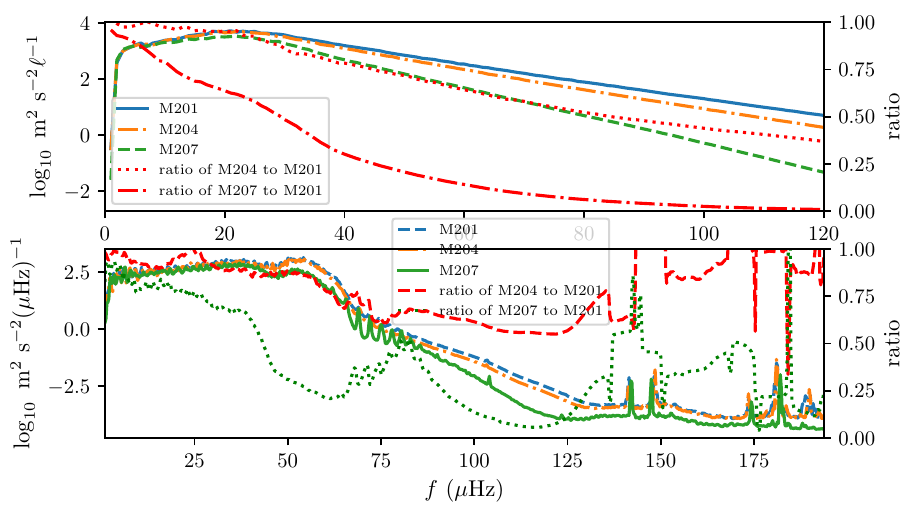}
  \caption{Power of radial velocity of M201, M204, and M207, as well as the power ratio of M204 and M207 to M201 at 19 $M_\odot$ as functions of spherical harmonic degree $l$ (top) and as functions of frequency (bottom).}
  \lFig{0kvs100k-damping}
\end{figure}
\begin{figure}
  \includegraphics[width=\columnwidth]{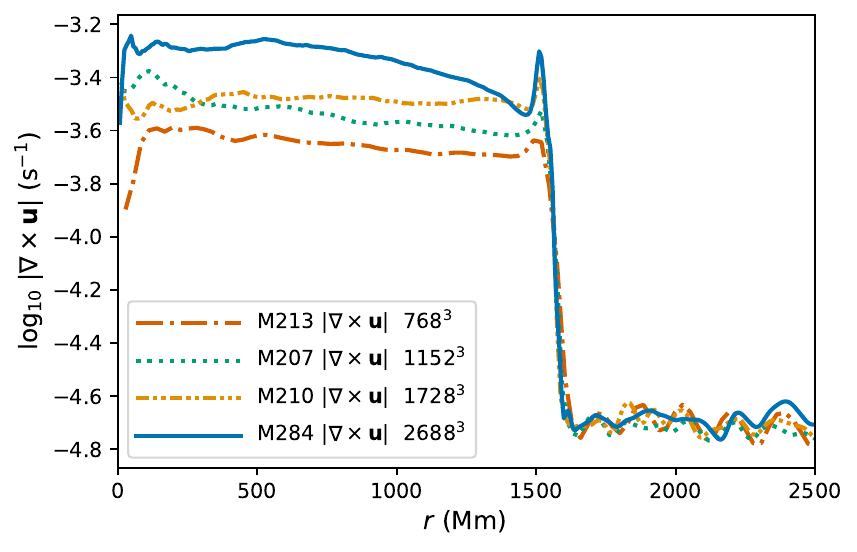}
  \caption{Vorticity of simulations with luminosity and radiative diffusion enhanced by 1000: M213 ($768^3$), M207 ($1152^3$), M210 ($1728^3$), M284 ($2688^3$) at 1518 h.  Although the vorticity increases with grid resolution inside the convection zone, it does not do so in the stably stratified envelope.  This behavior has consequences for our ability to estimate gravity-wave-based mixing rates in the envelope using simulations with only modest grid resolution.}
  \lFig{vortRes}
% This figure is produced by gradM200.ipynb.
\end{figure}

\section{On the long-term evolution}
\lSect{long-term}
The entrainment rate implied from linear growth of the entrained mass is too large to be compatible with the stellar
model and observational properties (\Sect{entrainment}). Similar to the argument in section 3 of \paperone, if we assume that this entrainment rate applies for the entire
\natlog{6.91}{6} \yr\ main sequence lifetime of a 25 \Msun\ star, a total entrainment of 630 \Msun\  would be implied. This indicates that the entrainment we
extrapolate cannot persist for even a fraction of the main sequence lifetime before the star goes through significant evolutionary changes. A motivation for the present
work is to investigate whether or not including radiation pressure and radiative diffusion can result in entrainment that is more consistent with the main sequence
stage of the stellar model. We have seen in \Sect{quasi-steady} above that this additional physics causes the entrainment to decrease by only about $30\%$ .
However, the linear growth of the entrained mass, the motion of the BV frequency peak, and the overshoot and undershoot of fluxes at the CB (\Fig{flux}) suggest that the simulated star
is still in the process of thermal adjustment. Nevertheless, the velocity distribution in both the convective core and the radiative envelope has reached
a dynamical equilibrium.  We would like to establish whether or not continued entrainment and motion of the CB outward might alter the
character of the flow in such a way that the entrainment rate might slowly diminish. This possibility is suggested by the recent work of
\cite{anders:21} investigating the long-term secular changes driven by thermal adjustment in a simplified Boussinesq, plane parallel, penetrative convection context.

Our explicit numerical technique in \code{PPMstar} requires us to explicitly follow sound wave signals in the low Mach number stellar flow.  We have seen in \paperone\ and also here in \Fig{entrainment} that we can overcome this limitation by appealing to empirically observed scaling laws.  By enhancing the luminosity and the radiative diffusion by a common factor $X$, we speed up the evolution by a similar factor (actually slightly larger than $X$, as we will discuss later).  In \paperone\ we saw that under these circumstances the velocities in the convection zone are enhanced by the factor $X^{1/3}$. If this enhancement of the velocities leaves them still at low Mach numbers, we do not expect the
character of the flow to change significantly. As a rule of thumb, we might attempt to hold the resulting Mach numbers below 0.1, for which compressibility
effects should be roughly of $1\%$ importance. A possible consideration is that we might raise velocities of wave motions in the stably stratified envelope to the level that either makes the waves break or that causes pressure to become an important restoring force influencing their dispersion relation. No wave breaking is observed in the stable envelope  in the visualizations of any of our flows. To validate this technique for speeding up the evolution of our flows, we can generate a series of simulations at modest grid resolution that have different enhancement factors $X$ and that can be compared over at least an initial time interval of a reasonable length, long enough to go through a noticeable re-adjustment of thermal structure.

\subsection{Key properties of the long-term evolution}
\lSect{Euler}
We have performed such a series of simulations for the 25 \Msun\ star which have enhancement factors $X$ = 1000, 3162, and 10000.  These
all have a grid resolution of $896^3$ cells, and all were run for relatively long periods of 507, 1189, and 1054 days for the star.  For the case of largest $X$, this
time duration is comparable to the thermal timescale of the simulated part of the 25 $\Msun$ star, namely $GM^2/2RL\approx 1000 \unitstyle{d}$, where $R=2500\ \Mm$ and $L=10000L_*$.  This should be sufficient for the flow to relax to a state much closer to thermal equilibrium.

In the top panel of \Fig{M252N2}, we show the outward movement of the BV frequency peak. This peak marks the location within the radial entropy profile where the gradient is largest.  This is also the location of the sudden jump in $f_{\rm V}$, the fractional volume of the stably stratified envelope gas.
It is evident that the outward motion of the CB is continually slowing down as this simulation proceeds.  The CB is still moving at the last time shown, but clearly it has slowed considerably.

Looking at \Fig{M252N2}, we notice that as the outward motion of the CB slows, there is an increasingly large region inside the CB (for time 17188 \hour\ between $r = 1600$ and 1750 \Mm) where the BV
frequency rises in the absence of any substantial contribution from the composition gradient.  This feature of the later flow structures causes the convection to be reduced in intensity
without causing additional entrainment.  It would seem that this is a necessary feature for the entrainment rate to be diminished. The positive entropy gradient that is established in the growing penetration region between the SB and the CB, results from a balance between convective mixing of entropy which tends to reduce this gradient, and the small region of negative gradient of the radiative diffusion flux, shown in \Fig{M252grad}, which tends to build up the gradient. There is no corresponding mechanism to counteract the convective mixing of the composition, because the negative radiative diffusion flux gradient deposits entropy and has no effect upon the gas composition. Hence we see that $f_{\rm V}$ is efficiently mixed, even in the penetration region.
\begin{figure}
  \includegraphics[width=\columnwidth]{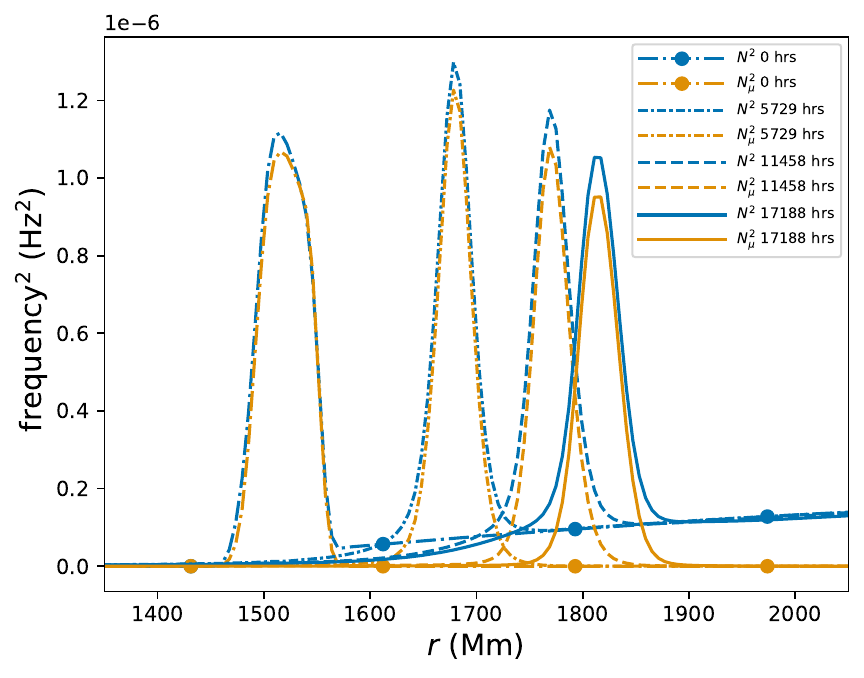}
  \includegraphics[width=\columnwidth]{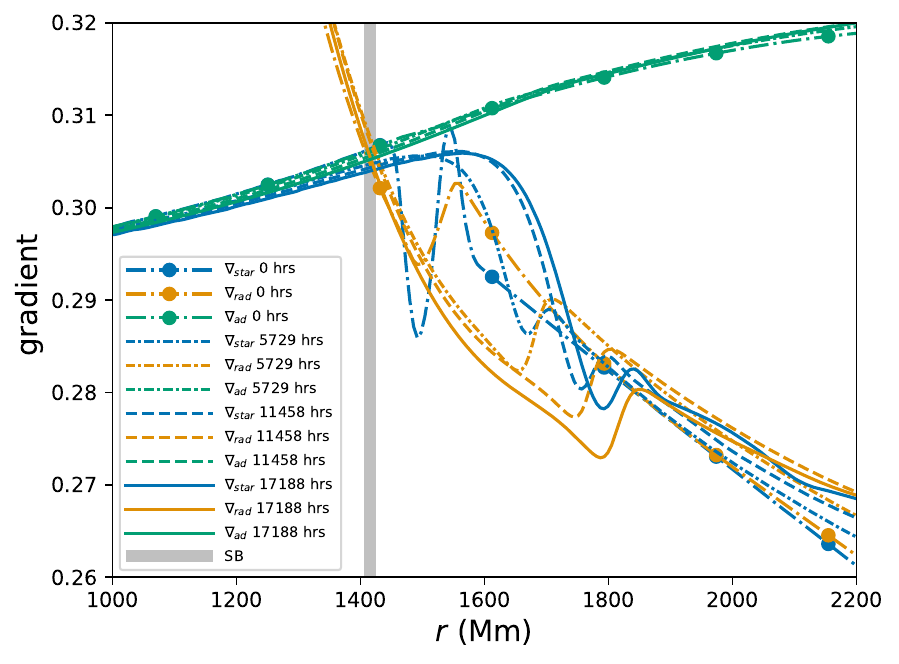}
  \caption{Top: $N^2$ and its compositional component (see \Eq{BV}) for 0, 5729, 11459, 17188 h. Bottom: radiative gradient, adiabatic gradient and actual temperature gradient for M252 (10000x) for the same dumps. The location of the SB is denoted by the thick vertical line around 1415 \Mm\ which does not move much during the simulation. All profiles are computed from averages over 100 dumps.}
  \lFig{M252N2}
\end{figure}

In \Fig{M252N2}, we plot the radiative gradient $$\nabla_\mathrm{rad}\equiv 3\kappa Lp/(16\pi acG mT^4)\ ,$$ the actual gradient $\nabla_{\rm star}$, and adiabatic gradient $\nabla_\mathrm{ad}$.
The radiative gradient is defined as the gradient required so that all the luminosity is carried outward by radiative diffusion.
The location, at roughly 1400 \Mm, of the SB, where $\nabla_\mathrm{ad}=\nabla_\mathrm{rad}$, does not change much during the course of the simulation.
The actual gradient is strictly adiabatic inside the SB at $t=0$ by design via initialization. When the convection is fully developed, the actual gradient becomes slightly super-adiabatic inside 1000 \Mm\ and slightly sub-adiabatic above 1000 \Mm\ (\Fig{supad}) and gradually approaches the radiative gradient above the SB, as seen in \Fig{M252N2}. The outward motion of the CB noted earlier slows down, which is also shown by the change of the actual gradient with time. The penetration region, where the convective flux is negative above the SB, ends at 1850 \Mm\, where the actual gradient starts to follow the radiative gradient, and the full luminosity is then carried outward by radiative diffusion alone (\Fig{M252grad}).
\begin{figure}
  \includegraphics[width=\columnwidth]{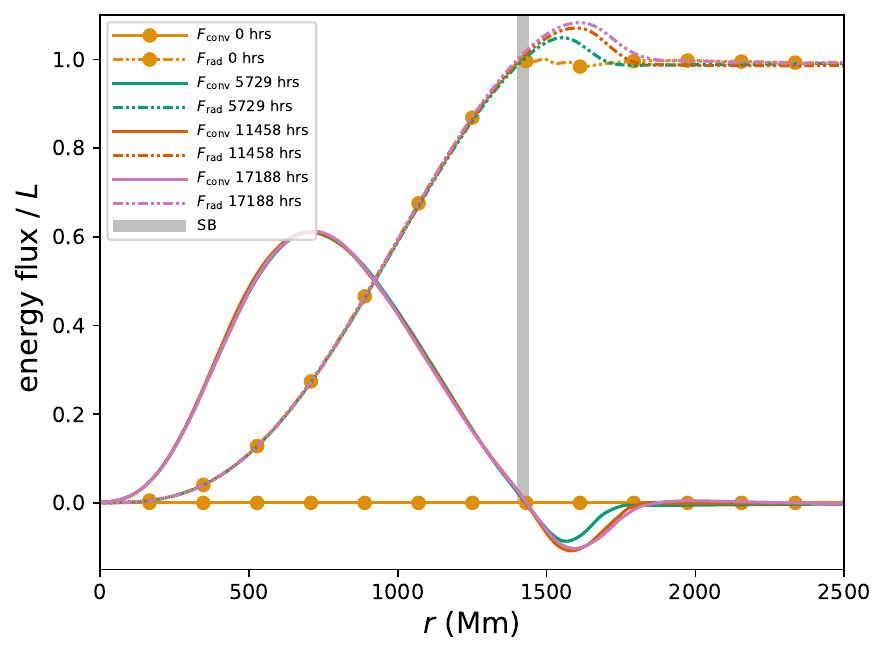}
  \caption{Total radiative and convective energy fluxes (normalized by the luminosity $L$) for M252 (10000x) at 0, 5729, 11459, 17188 h, computed from averages over 100 dumps and 120 \Mm.}
  \lFig{M252grad}
\end{figure}
% produced by gradM250.ipynb

\subsection{The governing equations}
\lSect{Euler}
Similar to \cite{anders:21,roxburgh1989integral,arnett2015beyond} (see also \cite{korre2021dynamics}), we attempt to model the convection zone by reducing the full set of hydrodynamic equations to 1-D with reasonable assumptions.
The governing hydrodynamics equations are the following:
\begin{eqnarray}
        \frac{\partial \rho}{\partial t} + \nabla\cdot(\rho \bm{u}) &=& 0   \lEq{conv_mass} \\
        \rho \frac{\partial \bm{u}}{\partial t}+\rho (\bm{u}\cdot\nabla) \bm{u} &=&
         -\nabla{p} + \rho \bm{g}  \lEq{conv_mom} \\
       T \frac{\partial}{\partial t}(\rho S) +  T \nabla\cdot(\rho S \bm{u}) &=& \epsilon \rho - \nabla\cdot\bm{F}   \lEq{conv_ent} 
\end{eqnarray}
where $S$ is the specific entropy, $\epsilon$ the rate of nuclear energy generation per unit mass, $\bm{F}$ the heat flux vector by radiative diffusion. These equations (\Eq{conv_mass} - \Eq{conv_ent}) are equivalent to the Euler equations in conservation form solved by \code{PPMstar}.
Taking the dot product of the equation for the conservation of momentum, \Eq{conv_mom}, with the velocity results in the equation for kinetic energy
\begin{eqnarray}
       \frac{\partial}{\partial t}(\frac{1}{2}\rho u^2)+\nabla\cdot(\frac{1}{2}\rho u^2\bm{u}) &=& - (\bm{u}\cdot\nabla)p + \rho \bm{u}\cdot\bm{g} \ .\lEq{ke_eqn} 
\end{eqnarray}

Without any assumption so far, we integrate the
kinetic energy density over a thin spherical shell between radius $r$ and $r+dr$ and determine its rate of change in time,
{\small
\begin{eqnarray}
\frac{\partial }{\partial t} \underset{(r,r+dr)}{\iiint} \frac{1}{2} \rho u^2 dV = & \underset{(r,r+dr)}{\iiint} (\rho \bm{u}\cdot\bm{g} - (\bm{u}\cdot\nabla)p ) dV \nonumber \\
& - \underset{(r,r+dr)}{\iiint} \nabla\cdot(\frac{1}{2}\rho u^2\bm{u})  dV 
\lEq{full-ke-eq}
\end{eqnarray}
}
where the kinetic energy equation \Eq{ke_eqn} is applied.
We apply the divergence theorem to the second term on the right-hand side and then approximate the resulting difference of surface integrals at $r$ and $r+dr$ with a differential, and finally approximate other volume integrals by surface integral multiplied by the shell thickness $dr$ to get 
{\small
 \begin{eqnarray}
 \frac{\partial}{\partial t}(\overline{\frac{1}{2}\rho u^2}4\pi r^2) &=
         \overline{(\rho_1 \bm{u}\cdot\bm{g} - (\bm{u}\cdot\nabla)p_1)}4\pi r^2  \nonumber \\
       & \quad -\frac{\partial}{\partial r}(\overline{\frac{1}{2}\rho u^2 u_r}4 \pi r^2) - 4\pi r^2 \overline{\Phi} 
         \lEq{kin-eqn} 
 \end{eqnarray}
}
as $dr\to 0$, where $p=p_0+p_1$, $\rho=\rho_0+\rho_1$,$\nabla p_0 =\rho_0 \bm{g}$. Note that subscript 0 denotes the initial hydrostatic state (or base state) quantities, subscripts 1 denote deviation from the initial hydrostatic state, $u_r$ is the radial velocity, $\Phi$ is the local dissipation rate of kinetic energy into heat, and the overbars represent averages over the $4\pi$ sphere at the radius $r$. Note that we assume that the viscosity does not enter directly, but only enters through the kinetic energy dissipation source term and entropy source term.  We make this assumption because the viscosity of the stellar gas is several orders of magnitude smaller than the thermal diffusivity, $k/(\rho c_p)$, where $c_p$ is the specific heat under constant pressure. (The Prandtl number is $Pr \sim 10^{-6}$ for the stellar interior conditions considered here.) Still, the viscosity is effective in dissipating the motions in the convection zone, while not dissipating motions elsewhere.  The reason for this effectiveness of a truly tiny viscosity is that the convection zone is fully turbulent. The turbulent cascade brings the motions there down to the tiny scales where the viscosity can act on them very efficiently to dissipate them into heat.  We will discuss how to determine this dissipation rate in \Sect{KE}.

We treat the entropy equation, given below, in a similar fashion to the kinetic energy equation:
{\small
\begin{eqnarray}
      \frac{\partial}{\partial t}(\overline{\rho S}4\pi r^2) &=
      4 \pi r^2 \overline{\dfrac{\Phi}{T}}  +\overline{\frac{1}{T}\nabla\cdot(\bm{\Gamma} -\bm{F})}4\pi r^2   \nonumber \\
     & \quad - \dfrac{\partial(\overline{\rho S u_r}4\pi r^2)}{\partial r}
     \lEq{ent-eqn}
\end{eqnarray}
}   
Here, for our convenience, we have defined $\bm{\Gamma}$ as the energy flux vector whose divergence gives us the nuclear energy generation rate, $\epsilon \rho$: 
$\nabla\cdot\bm{\Gamma} = \epsilon \rho$. \Eq{ent-eqn} and \Eq{kin-eqn} are the entropy equation and kinetic energy equation for each spherical shell.

\subsection{Reduced equations for kinetic energy and entropy}
\lSect{KE}
\subsubsection{Reduced kinetic energy equation}
\begin{figure}
  \includegraphics[width=\columnwidth]{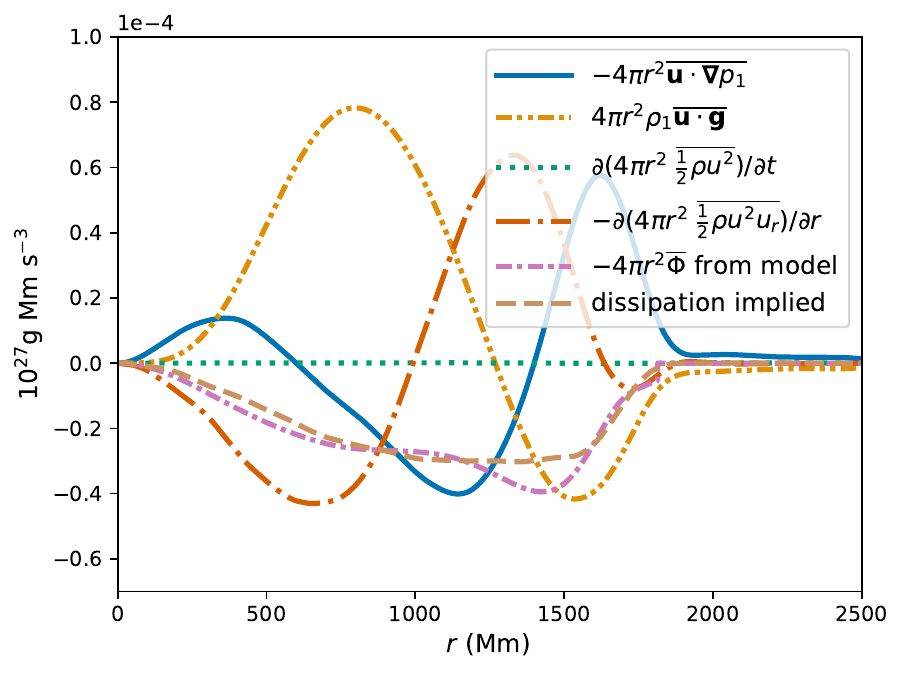}
  \includegraphics[width=\columnwidth]{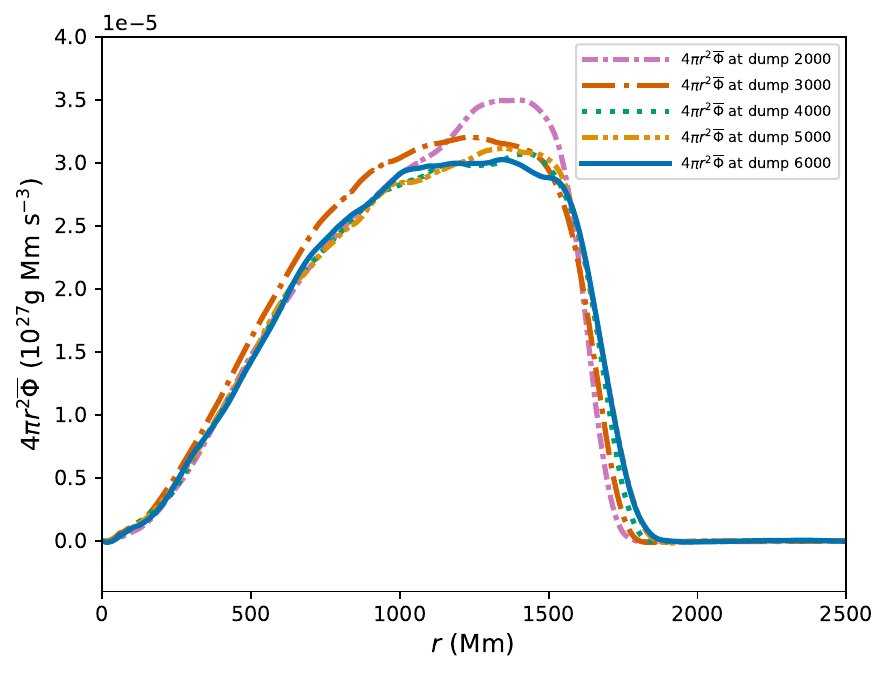}
  \caption{Top: work by pressure gradient and gravity field per unit time per unit radial distance, rate of change in kinetic energy per unit radial distance, radial derivative of total kinetic energy flux, dissipation derived from the turbulent dissipation model, implied dissipation rate per unit radial distance, averaged over 399 dumps ($\sim$ 1145 h) centered at 17188 h of M252; bottom: time sequence of measured turbulent kinetic energy dissipation of M252 at 5729, 8594, 11459, 14323, 17188 h.}
  \lFig{KE_M252}
\end{figure}
\begin{figure}
  \includegraphics[width=\columnwidth]{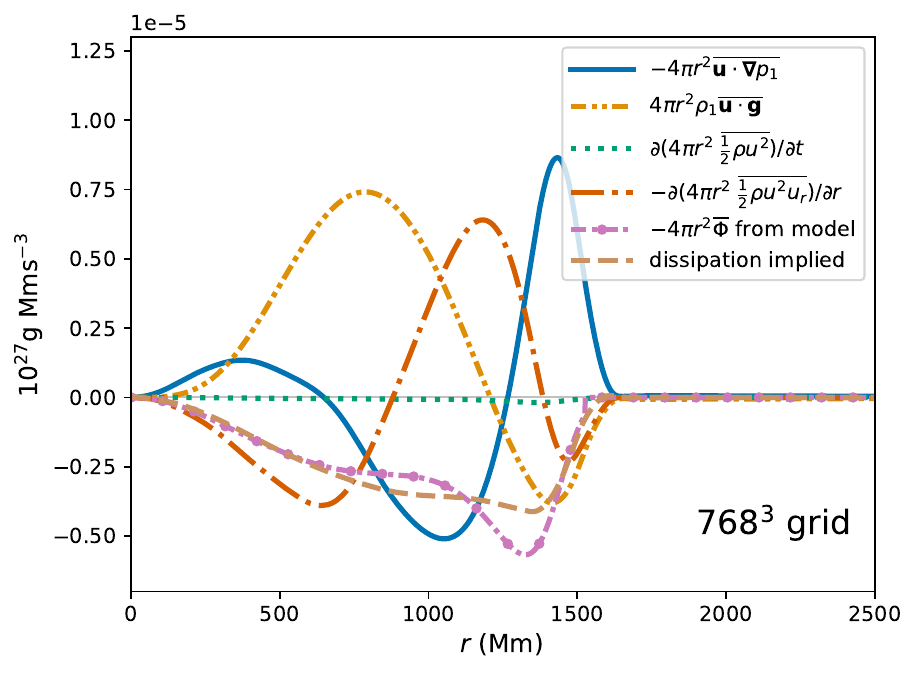}
  \includegraphics[width=\columnwidth]{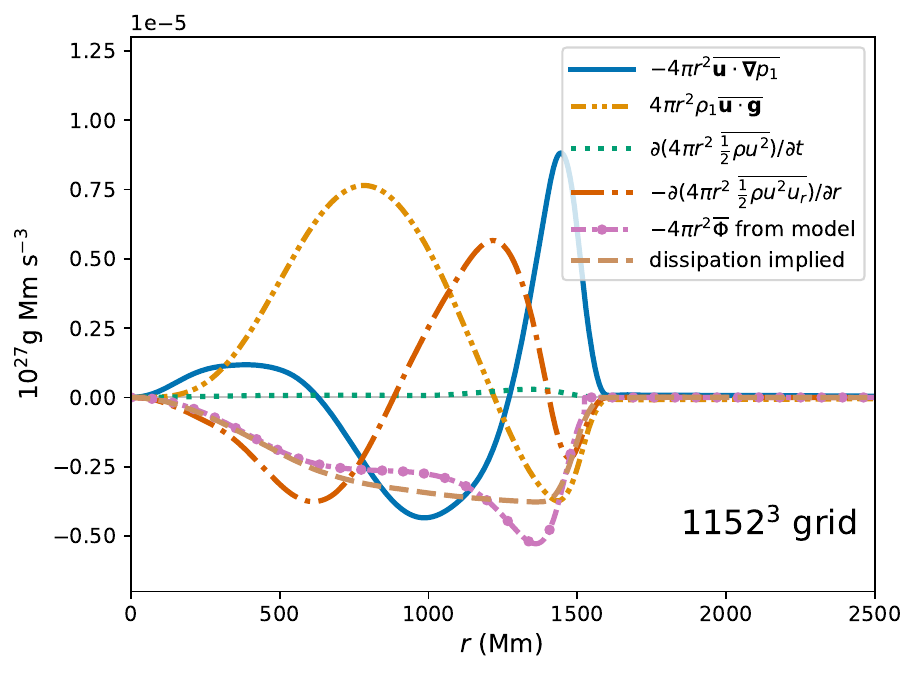}
  \includegraphics[width=\columnwidth]{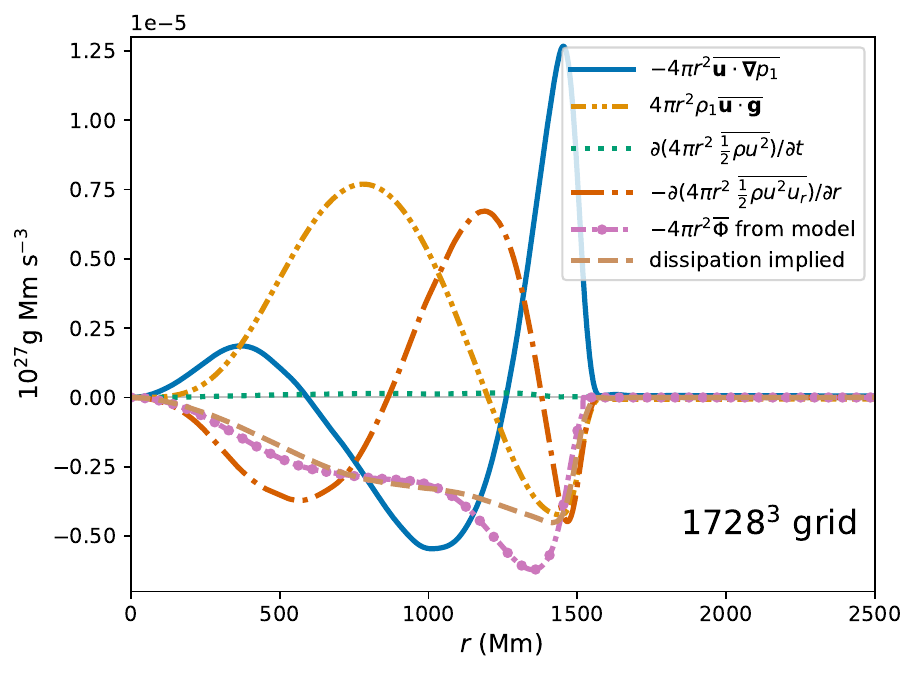}
  \caption{Work of pressure/gravity per unit time per unit radial distance, rate of change in kinetic energy per unit radial distance, radial derivative for the total kinetic energy flux, dissipation implied by the kinetic energy equation, dissipation from turbulence model for three resolutions of 1000x simulations (top to bottom: M213, M207, M210) at 1576.2 h averaged over 401 dumps ($\sim$ 286 h). Inside the SB at about 1400 \Mm\ there is little dependence of these values on grid resolution.  This means that we can get a good measurement of the implied turbulent kinetic energy dissipation rate using only a modest grid.}
  \lFig{KE_resolution}
\end{figure}
In our simulations, the gravity is static and determined by the base hydrostatic state. Therefore, the radial component of the base pressure gradient cancels out with the gravitational acceleration by design $\nabla p_0 =\rho_0 \bm{g}$. The gradient of the pressure perturbation, however, is not purely radial. Local high pressures can result in expansion in all directions. Hence, the pressure gradient term in the kinetic energy equation, which from the dot product evaluates to a scalar quantity, gives the work done by pressure per unit time per unit volume. The contribution of the horizontal components of the pressure gradient force $u \cdot \nabla p_1$ is not negligible, (\Fig{KE_M252}). In particular, the peak in $u \cdot \nabla p_1$ near the CB comes mostly from the horizontal component of the pressure perturbation gradient. In this region rising gas hitting the CB causes local high pressure, and the resulting flows are turned horizontal with large $u_\mathrm{h} \cdot \nabla_h p_1$.
Thus the pressure gradient force term is significant even in low Mach number flows and cannot be found in a 1-D computation except through a model, because of the 3-D nature of convection.

The \code{PPMstar} code solves the inviscid compressible fluid dynamics equations, and physical viscosity is not included.  This is reasonable, because the viscosity of stellar gas is truly miniscule.  However, in the convective core the convection is turbulent. Turbulent dissipation of kinetic energy is important in the convection zone.  This dissipation occurs via the turbulent cascade, which excites progressively smaller scales of motion until the viscous dissipation scale is finally reached.  In our simulations, this dissipation is carried out by numerical truncation error terms, some of which act like viscosity, but with different dependence upon the spatial scale of the motion, see \cite{Porter_Woodward_1994}.  The effectiveness of numerical methods like PPM in simulating turbulent flows in this fashion has been discussed at length and in detail, with many examples, in \cite{grinstein2007implicit, sytine2000convergence}.  There has been much work on modelling and theories for turbulent dissipation for stellar convection \citep[for example ][]{zahn1989tidal,porter1998inertial,woodward2006model,arnett2008turbulent}.
From the averaged kinetic energy equation \Eq{kin-eqn}, the dissipation term can be deduced from the rest of the other terms,
\begin{eqnarray}
  \begin{split}
-4\pi r^2\overline{\Phi} =& \frac{\partial}{\partial t}(\overline{\frac{1}{2}\rho u^2}4\pi r^2) 
 - \overline{(\rho_1 \bm{u}\cdot \bm{g} - \bm{u}\cdot\nabla p_1)}4\pi r^2   \\ 
 & +\frac{\partial}{\partial r}(\overline{\frac{1}{2}\rho u^2 u_r}4 \pi r^2)\ . \lEq{implied_dissipation}    
\end{split}
\end{eqnarray}

\cite{woodward2006model} estimates the turbulent dissipation as a function of density, and turbulent kinetic energy density for homogeneous, isotropic turbulence, 
\begin{eqnarray}
\frac{\partial E_\mathrm{turb}}{\partial t}=-A_{0}\frac{1}{L_0}\sqrt{\frac{2}{\rho}}E_\mathrm{turb}^{3/2} \lEq{tur_model}
\end{eqnarray}
where $L_0$ is the integral length scale which is the scale containing most of the kinetic energy, a dimensionless parameter $A_0 = 0.51$, $E_\mathrm{turb} =\frac{1}{2}\rho u^2$, $u$ is the turbulent velocity. By inserting the spherical averages of density and velocity magnitude of M252 in \Eq{tur_model} and using 1500 \Mm\ here empirically as the spatial scale that contains most of the kinetic energy, we get an estimate of turbulent dissipation from the model.

\Fig{KE_M252} presents the the terms in the kinetic energy equation, including the dissipation rate implied by the simulation from assuming that all the measured terms plus this dissipation must add to zero, and it also shows the dissipation rate derived using the turbulence model.
The core convection is not truly homogeneous, isotropic turbulence. However, its implied dissipation rate according to \Eq{implied_dissipation} agrees very well with the turbulent dissipation model.
The same model for turbulent dissipation with a different factor has been reported in \cite{frisch_1995} and \cite{Arnett2009}.
The agreement between the turbulent dissipation model \Eq{tur_model} and the dissipation rate indirectly measured from the simulation is striking. Note that the turbulent dissipation model does not apply above the CB, where, by our definition of the CB, the net convective entropy flux becomes essentially zero and any motions are no longer turbulent. We therefore do not apply the turbulent dissipation model at the CB and beyond. The dissipation in the convection zone is a result of the turbulent cascade only. This is confirmed by the dissipation from the simulation decreasing smoothly to zero at around 1835 \Mm\ in \Fig{KE_M252}. The dissipation of kinetic energy implied by the simulation is negligible in the radiative envelope.

For a disturbance of a fixed wavelength, the effective viscosity of the \code{PPMstar} method scales as the cube of the grid cell size $\Delta x$ \citep{Porter_Woodward_1994}.  Therefore, each 1.5x grid refinement implies a decrease in the numerical viscosity at each wavelength by a factor of 3.375.  Nevertheless, the results plotted in \Fig{KE_resolution} show that the dissipation of kinetic energy in the convection zone is independent of the grid resolution for grids equal to or finer than $768^3$ for our \code{PPMstar} code.  This apparent contradiction can be explained by the action of the turbulent cascade, in the effective absence of viscosity, transporting kinetic energy from larger to smaller scales at a rate that is independent of scale.  This self-similarity of the turbulent flow is the basis of the \cite{kolmogorov1941local} argument for the power-law spectrum shown in \Fig{KEpower}. When the kinetic energy reaches scales small enough that the viscosity becomes important, this energy is damped and transformed into heat.  In the star, this occurs at tiny length scales much smaller than the width of a single cell on any of our computational grids.  In our simulations, this occurs on length scales of a few grid cell widths.  In \Fig{KEpower}, we see that the damping, which causes the power at a given wavelength to fall below the Kolmogorov trend, sets in at shorter wavelengths as the grid is refined.

\subsubsection{Verification of turbulent dissipation measurement}

\begin{figure}
  \includegraphics[width=\columnwidth]{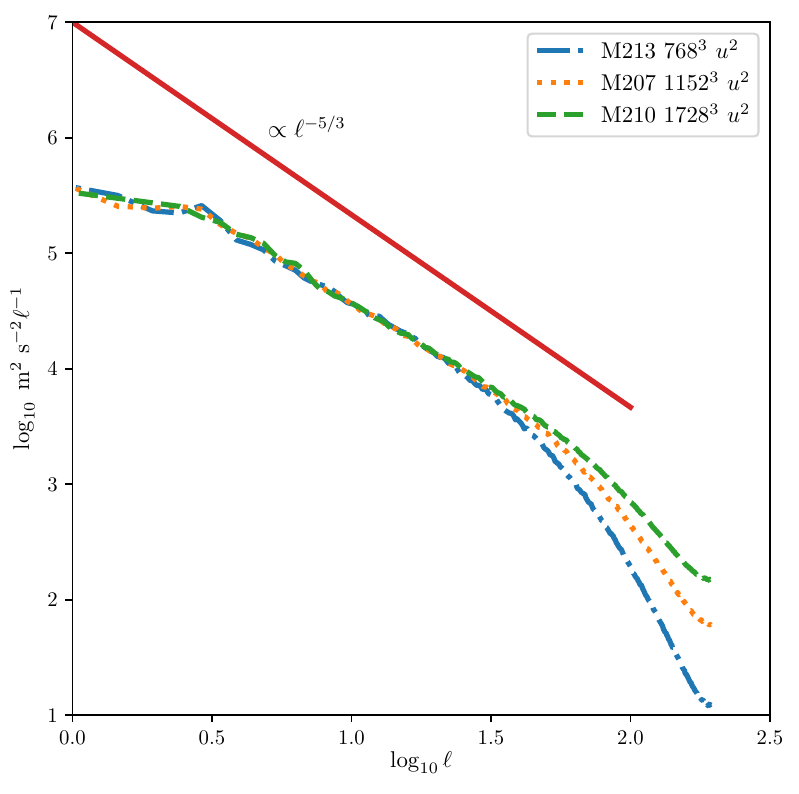}
  \caption{Velocity power spectra at radius 1000 \Mm\ for 3 runs, at grid resolutions of $768^3$, $1152^3$, and $1728^3$ cells, averaged over 573 hours (800 dumps) centered at \unit{1576}{\hour}. { These spectra where calculated using the filtered briquette data outputs of \code{PPMstar} which has four times less resolution in each dimension than the actual grid.}}
  \lFig{KEpower}
% plotted by Ekin_M216-M218.ipynb
\end{figure}

Three simulations are performed, which restarted from a late time (dynamical equilibrium already established) of the 1000x heating and 1000x radiative diffusion cases with 3 resolutions (M213, M207 and M210). Volume heating and radiative diffusion are turned off from the beginning of these three new runs. The intent is to measure the decay rate of the kinetic energy in the convective core, which should be the same as the turbulence dissipation rate.
The kinetic energy per unit volume is plotted about every 8.5 hours in \Fig{Ekin_M216-M218}.
Before the nuclear heating is removed, we have a slightly convectively unstable stratification.
The unstable stratification continues driving the convection for a short while before it is eliminated. Hence, the decay of kinetic energy is barely noticeable in the first couple of dumps.
The total decay rates of kinetic energy are estimated from the first 60 hours to be $20.5\%$, $17.8\%$ and
$19.3\%$ (from low to high resolution) of the luminosity. Again, we do not see kinetic energy dissipated in the stable envelope.
\begin{figure}
  \includegraphics[width=\columnwidth]{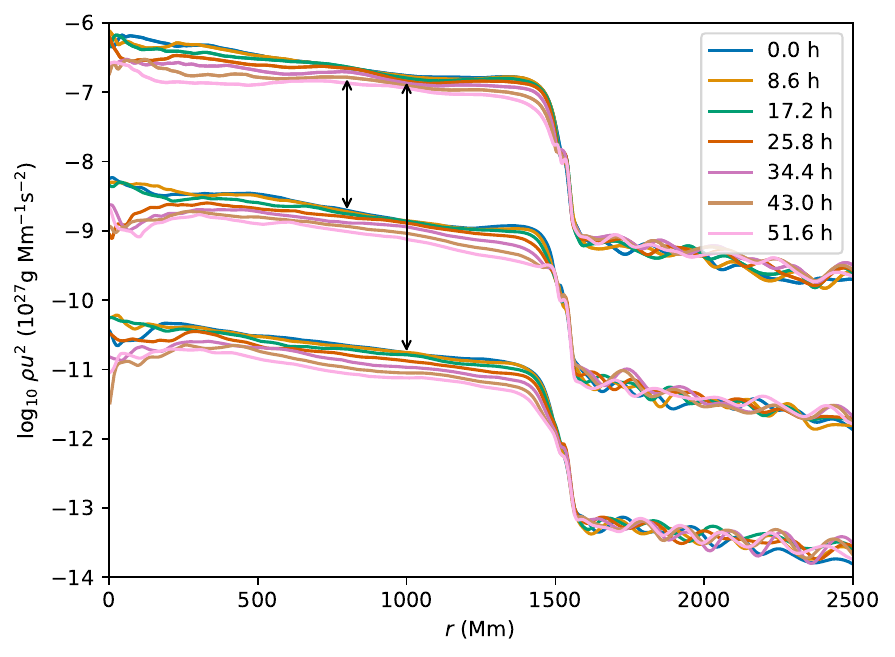}
  \caption{Time sequence of kinetic energy density every 8.5 hours for the rundown experiments of M210: $1728^3$, M207: $1152^3$ and M213: $768^3$ (upper, medium and lower groups of lines) where the medium and
lower resolutions have been translated downward by 2 and 4, for easy visual comparison; Note that essentially no dissipation is seen outside the convection zone.}
  \lFig{Ekin_M216-M218}
% plotted by Ekin_M216-M218.ipynb
\end{figure}

\subsubsection{Reduced entropy equation}
\begin{figure}
  \includegraphics[width=\columnwidth]{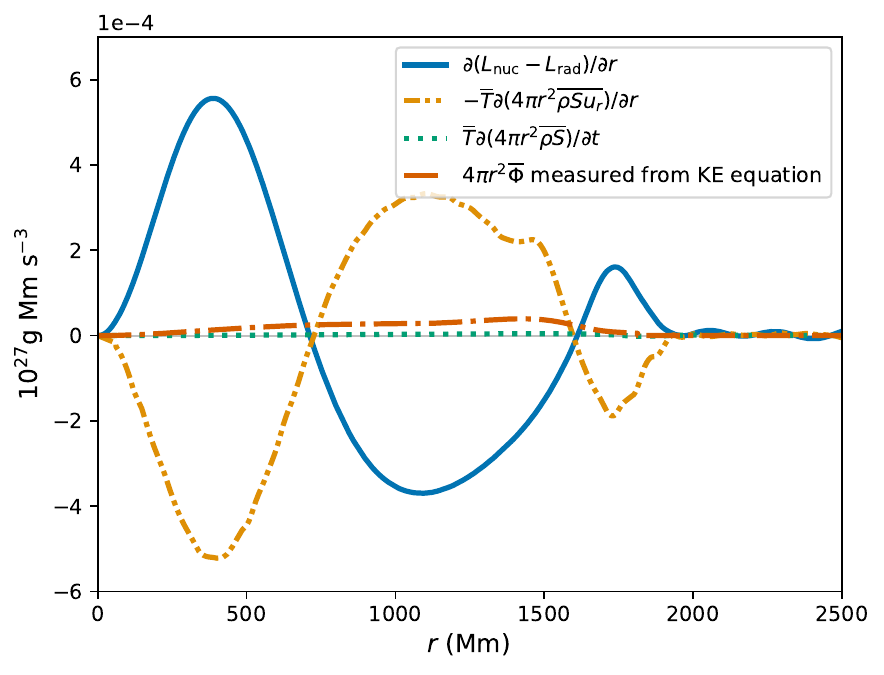}
  \caption{The gradient of energy flux from nuclear heating and radiative diffusion, gradient of the advective entropy flux multiplied by temperature, rate of change in entropy multiplied by temperature, dissipation of kinetic energy measured at 17188 h of run M252.  Note the small role played by turbulent kinetic energy dissipation relative to the other terms in the entropy equation.}
  \lFig{ENT_M252}
% produced by KE-ENT.ipynb
\end{figure}
Now we proceed to investigate the reduced entropy equation to see if it leads to useful 1-D modelling that has predictive power on whether the star is in equilibrium or how big the convective penetration region
should be. There is no approximation
in deriving \Eq{ent-eqn}. The entropy equation simply states that the rate of change of entropy in a spherical shell is the sum of turbulent dissipation of kinetic energy,
heating and cooling of nuclear burning and radiative diffusion, and the advective flux of entropy. To further simplify, we assume the radiative energy flux vector is a function of radius alone and is radially directed, which is not strictly true because the adiabatic motion will heat or cool fluid parcels, and then the heat flux can have a non-zero horizontal component. The second term on the right-hand side then becomes
\begin{equation}
  \overline{\frac{1}{T}\frac{\partial(4\pi r^2(\Gamma_r-F_r))}{\partial r}}\equiv \frac{1}{\overline{T}}\frac{\partial(L_\mathrm{nuc}-L_\mathrm{rad})}{\partial r}
  \lEq{Lrad-eqn}   
\end{equation}
This relation serves as a definition for $L_\mathrm{nuc}$ and $L_\mathrm{rad}$.

The terms in the reduced entropy equation
\begin{align}
  4 \pi \overline{T}\dfrac{\partial}{\partial t}(\overline{\rho S} r^2) &=
  4 \pi r^2 \overline{\Phi}  +\dfrac{\partial}{\partial r}  (L_\mathrm{nuc} - L_\mathrm{rad})  \nonumber \\
 &\quad - 4 \pi \overline{T} \dfrac{\partial(\overline{\rho S u_r} r^2)}{\partial r} \lEq{red_entropy_eqn}
\end{align}
are shown at a very late time in \Fig{ENT_M252} for our run M252.  By the time shown, namely 17188 hours, the time rate of change of entropy is nearly zero at all radii, and the convection zone has expanded considerably from its position at early times in \Fig{KE_resolution} or in the first panel of \Fig{M252image}. The time shown in \Fig{ENT_M252} matches that shown in \Fig{KE_M252}.  With this very long duration run at the luminosity enhancement factor of 10000, we have been able to bring both the kinetic energy equation (\Fig{KE_M252}) and the entropy equation (\Fig{ENT_M252}) simultaneously into very near equilibrium. 

\subsubsection{Accelerating stellar evolution by enhancing luminosity and radiative diffusion}
\lSect{acc}

A key challenge of core convection simulations is the large ratio between the thermal and convective time scales. We address this disparity by boosting the luminosity and reducing the opacity by a common factor $b$.  Because the thermal time scale scales with $b^{-1}$ and the convective, or dynamic time scale scales with $b^{-1/3}$,  the ratio of thermal to dynamic time scale becomes smaller with larger boost factors.  Boosting the luminosity therefore makes it computationally less costly to carry the simulation forward long enough to approach a dynamic and thermal equilibrium. For this reason luminosity boost factors of $10^5$ and $10^6$ are commonly found in the literature \citep[see, for example ][]{Andrassy:24,Edelmann:2019jh}. Restriction of the problem to just 2-D  \citep[see, for example ][]{Baraffe:23} or to plane-parallel geometry \citep[][]{anders:21} are other strategies to keep computational costs down. To capture the largest convective eddies in core convection requires simulating the entire core convection zone (in 3-D of course), as can clearly be seen from the results shown in \Fig{image}.  For our very long run M252, with a luminosity boost factor of $10000$, we are able to carry the simulation forward for a thermal timescale, which brings the convection flow into very near thermal and dynamical equilibrium.

\begin{figure}
  \includegraphics[width=\columnwidth]{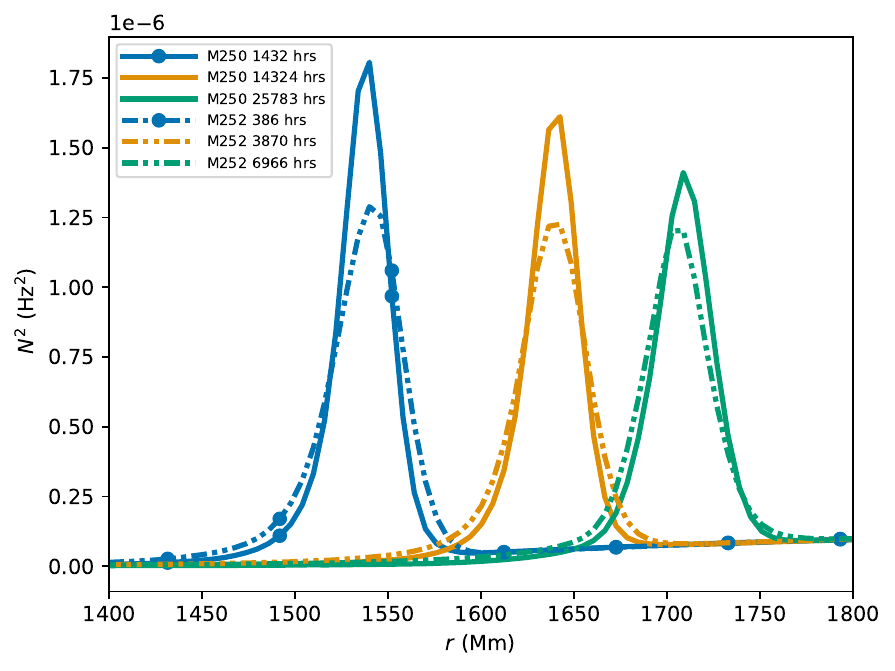}
  \includegraphics[width=\columnwidth]{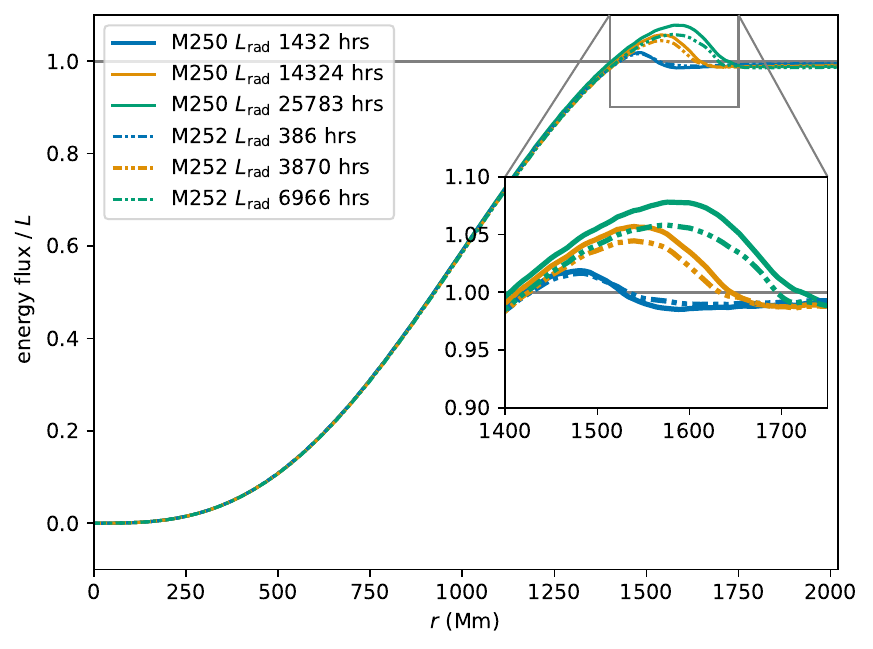}
  \caption{$N^2$ of M250 (3162x) and M252 (10000x) at three different times when they proceed to the same location; Bottom: total radiative heat flux of M252 (10000x) and M250 (3162x) normalized by their respective luminosities $L$.  Considering that $N^2$ depends upon the local entropy \emph{gradient}, it is remarkable how similar the results of these two runs are at these times, especially considering the more than 3 times greater computational cost of the M250 results.}
  \lFig{accel_scale}
\end{figure}
% produced by gradM250.ipynb

Comparing the vertical scales of \Fig{KE_resolution} and \Fig{KE_M252} suggests that the terms in the kinetic energy equation scale linearly with the boosting factor. The scaling of convective velocity with luminosity (\Fig{u-L}) and the turbulent dissipation model \Eq{tur_model} also imply that the turbulent dissipation scales linearly with the luminosity enhancement. Hence, the turbulent dissipation, $\bm{F}$, and $\bm{\Gamma}$ in \Eq{ent-eqn}, all scale linearly with the boosting factor $L/L_{*}$. The time rate of change in entropy is driven to become very small on the thermal time scale, so that the star is nearly thermally relaxed, as is clear in the late-time plot for our run M252 shown in \Fig{ENT_M252}.  Once the time rate of change of entropy is driven nearly to zero in this way, we have argued above that all the terms in the entropy equation except the convective entropy flux scale linearly with luminosity.  Because all these terms plus the convective entropy flux term then add to essentially zero, that flux term must also scale linearly with luminosity.  Therefore, we conclude that when the stratification is close to equilibrium, and the time rates of change for both kinetic energy and entropy nearly vanish, then all the other terms in the kinetic energy and entropy equations must scale linearly with the luminosity enhancement factor.  It is natural to hope that the rates of change with time of kinetic energy and entropy also scale linearly with luminosity enhancement, so that we can accelerate our simulations by boosting the luminosity and thermal conductivity by the same factor.

\begin{figure}
  \includegraphics[width=\columnwidth]{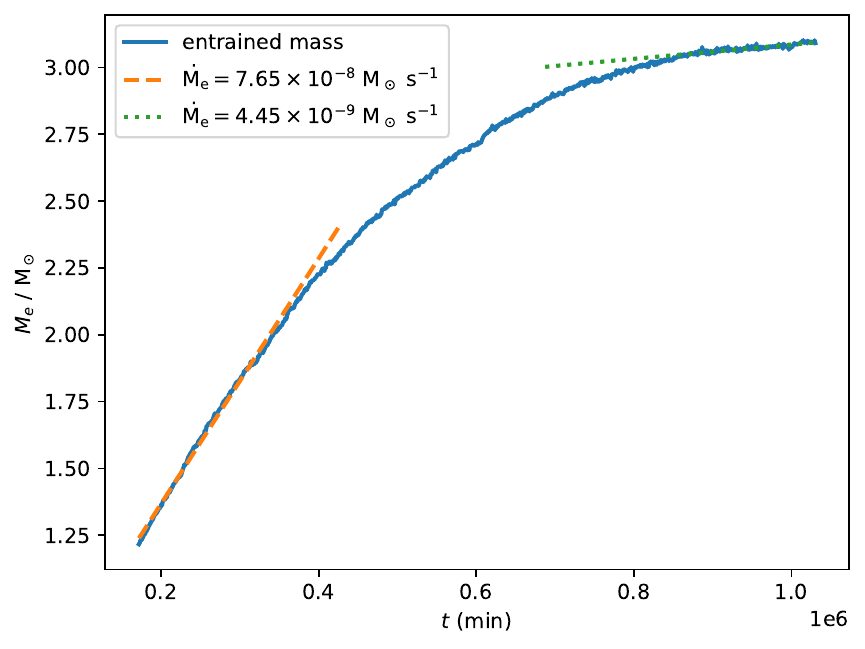}
  \caption{Entrained mass as a function of time of M252.  Over the course of this long simulation, the entrainment rate has dropped by about a factor of $\sim 17$, although it has not fallen to zero when the simulation was stopped. The entrainment rates are measured from 2865 to 5729 h, and from 14323 to 17188 h}
  \lFig{entrainmentM252}
\end{figure}
\begin{figure}
  \includegraphics[width=\columnwidth]{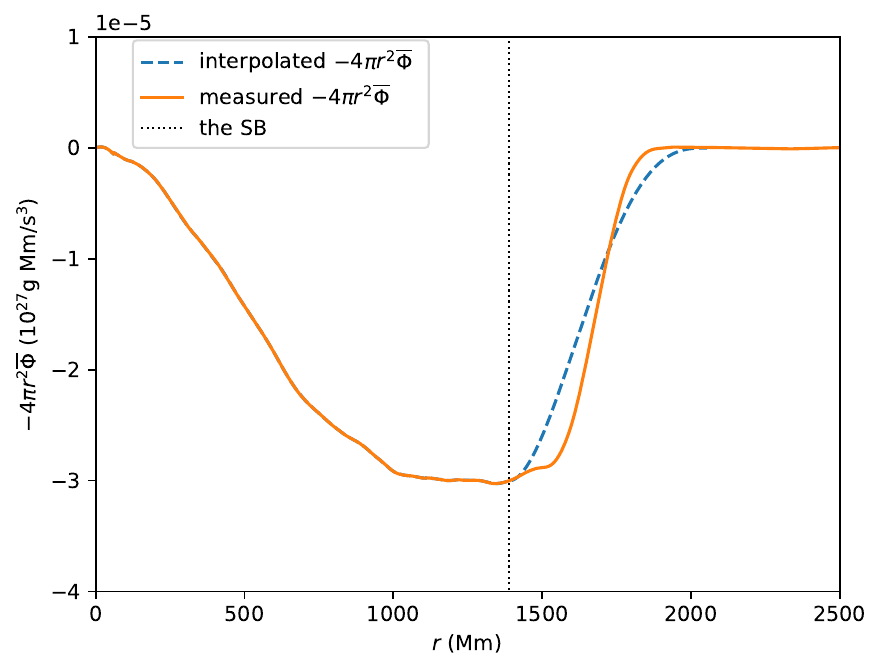}
  \includegraphics[width=\columnwidth]{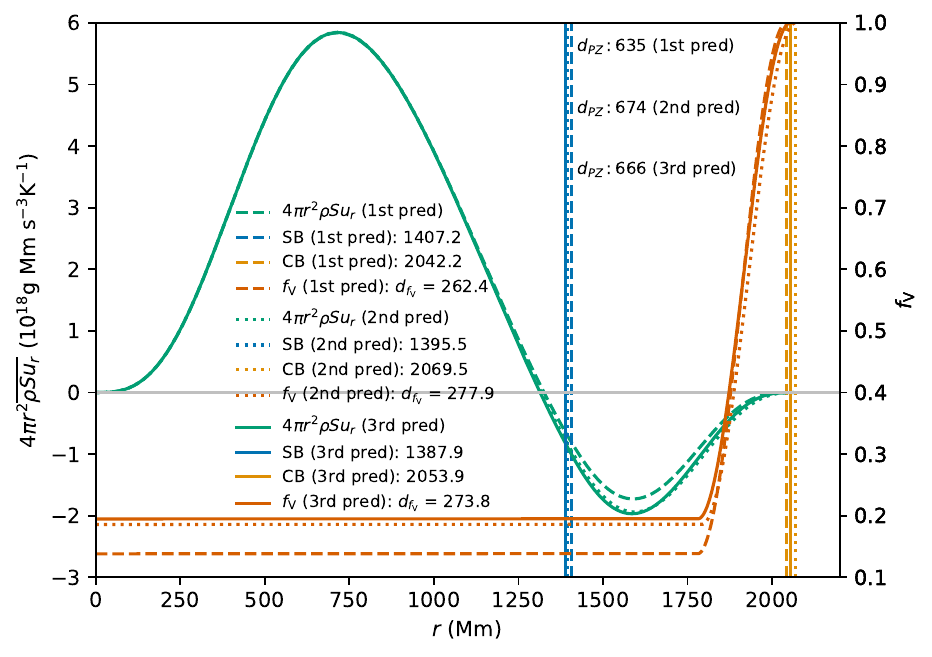}
  \caption{Top: dissipation measured from M252 at 17188 h and interpolated beyond the SB; bottom: entropy flux implied by the predicted hydrostatic equilibrium stratification using the measured dissipation inside the SB and the interpolated dissipation above the SB. The 1st prediction uses central density, entropy and $f_{\rm V}$ and turbulent dissipation below the SB at 5729 h, 2nd at 11459 h, 3rd at 17188 h.}
  \lFig{excessSfluxTur}
\end{figure}
\begin{figure}
  \includegraphics[width=\columnwidth]{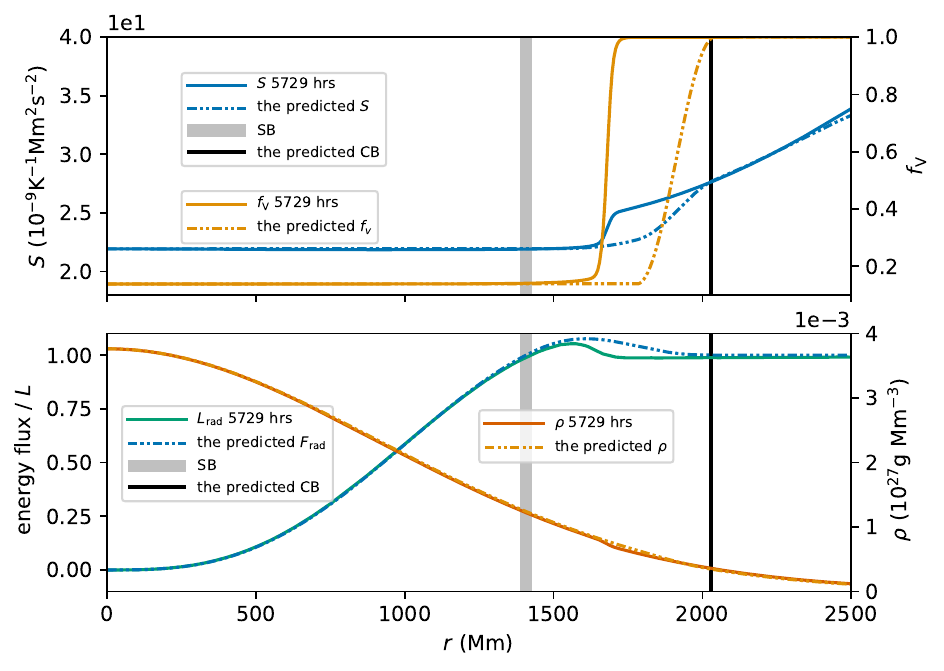}
  \includegraphics[width=\columnwidth]{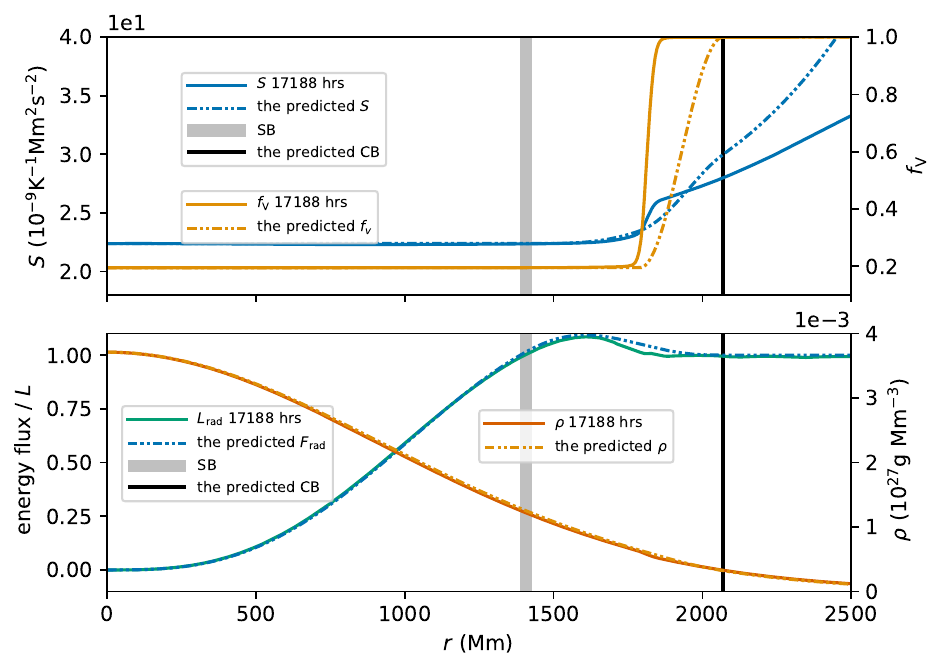}
  \caption{The thermal equilibria predicted by the 1-D method and the stratifications of M252 (10000x $L_*$ \& $k_*$) at 5729 h (top) and 6000 (17188 h, bottom).}
  \lFig{simVSpred}
\end{figure}
% produced by gradM250.ipynb
We have performed a series of very long simulations on grids of $896^3$ cells for luminosity enhancement factors of 1000, 3162, and 10000.  These are the final three runs listed in \Tab{run_tab}.  These long runs give us an opportunity to test the conjecture that boosting the luminosity and thermal conductivity by the same factor speeds up the approach to a single common equilibrium stratification by approximately that same boost factor.  In the bottom panel of \Fig{accel_scale} we show the radiative heat fluxes, normalized by the boosted luminosities, of the two higher luminosity runs at times proportional to the inverse of their boost factors. To the degree that these  curves agree, the conjecture is true.  In the top panel of \Fig{accel_scale}, we show the square of the BV frequencies for these two runs, plotted at three different times when these frequency peaks had moved to the same location in radius. These BV frequency curves are quite similar. The times when these two runs have their BV frequency peaks at the same location differ not strictly by the 3.162 factor by which their luminosities differ, but instead by a factor of about 3.6. For our two long runs, M250 and M252  at luminosity boost factors of $b=1000$ and $10000$, the Brunt-V\"ais\"al\"a frequency peaks  for the run with higher luminosity are not as high.  This means that the strong changes in entropy and composition at the convective boundary have somewhat gentler slopes in this case.

The lower BV frequency peaks in our higher luminosity simulation are consistent with a trend that we have noticed before in the thickness of the CB region, as measured by the fitted overshooting parameter $f$, scaling with luminosity to the 1/3 power \citep{woodward:19,Denissenkov:19}.  \cite{Baraffe:23} also report a dependence of the \emph{overshooting length} of convection in 2-D simulations on the 1/3 power of the luminosity.  More direct evidence of such a dependence of the thickness in 1-D averages of the convective boundary region, and also of the convective boundary location, has been recently reported by \cite{Andrassy:24}.  Earlier studies published by \cite{Baraffe2021} and by \cite{Kapyla:20} were inconclusive on this point.  \cite{Baraffe2021} used 2-D simulations and did not carry them out through a full thermal adjustment time.  Their results were consistent with our arguments here that higher luminosity boosts accelerate the outward motion of the convective boundary toward its equilibrium position.  \cite{Kapyla:20} did not find any significant luminosity dependence of their penetration depths, but those penetration regions were constrained in size by the nearby location of the boundary of their computational region.  Reviewers of the manuscript for this article encouraged us indirectly to consider that a model of the convective penetration region must involve some parameter that accounts for the dependence of the penetration depth on luminosity.  Our procedure described in the next section shows how we can model the equilibrium structure of the convection zone appropriately for large luminosity boosts, such as the factor 10000 used in our run M252 that is discussed above, and sketches how the procedure can be modified to incorporate a parameterized convective boundary thickness in future work.

\section{A 1-D model of  a convection zone with penetation that is in dynamic and thermal equilibrium}
\lSect{penetration}
We have shown in \Fig{KE_M252}, \Fig{ENT_M252} and \Fig{accel_scale} 
that our very long simulation, run M252, has come very close to a state of dynamic and thermal equilibrium.  At the latest times in this run, as are shown in \Fig{entrainmentM252}, the convective boundary is still moving outward but at a pace reduced by a factor of 17.  In its long evolution, this run provides us with sufficient information to extrapolate its approach to equilibrium and thus to approximate its ultimate equilibrium state.  Our analysis of the 1-D kinetic energy and entropy equations, \Eq{kin-eqn} 
and \Eq{red_entropy_eqn}, provides the context for this extrapolation procedure.  We note that the dynamic equilibrium expressed by the kinetic energy equation with vanishing time derivative (\Eq{implied_dissipation})    with the first term on the right set to zero, is established relatively rapidly.  As the region of penetrative convection is slowly extended, the kinetic energy dissipation rate,  $\Phi(r)$, changes hardly at all for r inside the SB, as is shown in \Fig{KE_M252}.  We can therefore use \Eq{implied_dissipation}  with zero time derivative to solve for  $\Phi(r)$ in this region.  This requires, of course, that we perform a short 3-D simulation, but our results show in   \Fig{Ekin_M216-M218}
that a modest grid of $768^3$ cells is sufficient.  We have remarked earlier that a 3-D simulation in the correct spherical symmetry is needed to determine the pressure gradient term in \Eq{implied_dissipation}, but if we were to have available a series of such short simulations for stars of different masses and evolutionary states, we would be able to find a very good approximation for $\Phi(r)$ inside the SB by means of interpolation.  We can also see from \Fig{KE_M252}, and the results in  \Fig{KE_resolution} as well, that we may extend  $\Phi(r)$ to the CB, where we know it must vanish, by using the unique quintic polynomial that assumes the known value of  $\Phi(r)$ at the SB, vanishes at the CB, and has vanishing first and second derivatives at the SB and CB.  This approximation assumes that we know the radius, $r_\mathrm{CB}$, of the CB.  We will make a guess at $r_\mathrm{CB}$ and improve it iteratively.

We now turn our attention to the entropy equation \Eq{red_entropy_eqn}. We will set the time derivative term to zero, since we seek an equilibrium state. As we have remarked earlier, we can use this equation to solve for the convective entropy flux term, $4\pi \frac{\partial (r^2 \overline{(\rho S u_\mathrm{r})})}{\partial r}$. Inside the SB, we have this term already from our short 3-D simulation, but especially at lower luminosities this term tends to exhibit much more fluctuations than the others in \Eq{red_entropy_eqn}, so that our short simulation might not have provided a good estimate. We will use equation \Eq{red_entropy_eqn} to solve for the convective entropy flux term in the entire convection zone, all the way out to the CB. We take the nuclear heating rate, the term $\frac{\partial L_\mathrm{nuc}}{\partial r}$, from our short simulation inside the SB and assume it to vanish outside that radius. Now, in order to use equation \Eq{red_entropy_eqn} to solve for the convective entropy flux term, we must extend the radiative diffusion flux, $L_{\mathrm{rad}}(r)$, which is known inside the SB, outward to the CB. A simple model for this flux is to use the unique quintic polynomial that assumes the values of $L_{\mathrm{rad}}(r)$ and its first two radial derivatives at the SB and that also assumes the value $L$, the total luminosity, at the CB, with its first two radial derivatives vanishing there. We have found that this continuation of $L_{\mathrm{rad}}(r)$ from SB to CB is appropriate to a very large luminosity boost factor, such as the value 10000 used in our run M252.

Once we have chosen the forms of the continuations of both $\Phi(r)$ and $L_\mathrm{rad}(r)$ from the SB to the CB, we seek the value of $r_\mathrm{CB}$ that results in a vanishing value there of the convective entropy flux, $\overline{(\rho S u_\mathrm{r})}$. In order to evaluate the convective flux at the CB, we integrate the convective entropy flux term, $4\pi r^2 \overline{(\rho S u_\mathrm{r})} / \partial r$, outward from the origin to the CB. We find that a unique value of $r_\mathrm{CB}$ results from the demand that the entropy flux $\overline{(\rho S u_\mathrm{r})}$ must vanish at the CB. We may then derive the entire stratification in the convection zone by demanding that the gas be hydrostatic and lie on the same adiabat as the gas at the origin up to the SB, and by demanding that it be hydrostatic and produce the prescribed extended $L_{\mathrm{rad}}(r)$ values from the SB to the CB. Such a projected equilibrium state for the case of our run M252 is shown in \Fig{excessSfluxTur}
for projections made at three different times during that simulation. These three projections are very closely the same. Projected equilibrium states made at two different times for run M252 are plotted against the simulation in  \Fig{simVSpred}.  The results shown in  \Fig{simVSpred} show that we have assumed in our projections that the composition jump, parameterized via
\begin{eqnarray}
   f_{\rm V}(r)= & \frac{1}{2}[1+\sin(\pi\frac{r-r_{\rm foot}}{r_{\rm CB}-r_{\rm foot}}-\frac{\pi}{2})](1-f_{\rm V}(r_{\rm SB})) \nonumber \\  
    &   + f_{\rm V}(r_{\rm SB})  \lEq{fvjump}
\end{eqnarray}
where $r_{\rm foot}<r<r_{\rm CB}$,
begins arbitrarily at the point where the radial derivative of the extended $L_{\mathrm{rad}}(r)$ has its most strongly negative value. The plots in \Fig{simVSpred} indicate that this is likely to be a mistake, because the projected composition jumps are much gentler and thicker than those in the simulations at both times shown. Instead, it appears that the composition jump, the jump in our variable $f_V$, should begin very close to the CB. Our results of the convergence study shown at an early time in     \Fig{Prad-1000x-K-L-res} indicate that the thickness of the composition jump is likely to be unresolved on our $896^3$ grid in run M252. The simulation produces fairly sharp jumps in entropy near the CB at both times shown in  \Fig{simVSpred}. A strong component of these jumps comes from the composition jump in $f_V$. However, the entropy jump at the later time shown has a gradual rise before it that is caused by a balance between local heating from a declining radiative flux and cooling by the action of penetrative convection, as we have remarked earlier. The thickness of the entropy jump at the CB in a case where there is no composition difference between the convection zone and the radiative envelope will therefore be determined solely by the thickness of the region where the radiative flux returns from its overshooting value to the total luminosity in the penetrative region. In the projected equilibrium models shown in  \Fig{excessSfluxTur} and  \Fig{simVSpred} we have essentially assumed that the thickness of this transition of $L_{\mathrm{rad}}(r)$ from the adiabatic value to $L$ is simply the width of the penetration region. This assumption is appropriate for high luminosities, as is the case for our run M252.

We argued earlier that all the terms in the kinetic energy and entropy equations should scale linearly with luminosity. Some of these terms clearly do scale this way, and in an equilibrium state the time derivative terms will vanish. However, this scaling cannot be precisely exhibited by all the terms in these equations, because this would result in impossible entropy structures in the equilibrium penetration regions. To see this, consider the gradual entropy increases in approaching the CB that are plotted for the projected equilibria in  \Fig{simVSpred}. At a high luminosity value, with also an equally enhanced value of the convective entropy flux in the penetration region, the convection will have no problem continuing despite the small but non-zero adverse entropy gradient in this region. However, as the luminosity is reduced, the weakening convection will be stopped by this small entropy gradient. At a lower luminosity, the entropy gradient in the penetration region would have to be reduced toward zero for the convection to exist there. This reduction in the entropy gradient would have to be accompanied by a change in the radial behavior of $L_{\mathrm{rad}}(r)$ in the penetration region.

We have performed a series of simulations at different luminosities all beginning with a projected equilibrium state for our 25 M$_\odot$ model star at an earlier time in its evolution, when there is no composition gradient. These simulations will be reported in a future article. They do show, however, that both the thickness and the radial location of the entropy jump at the CB change as anticipated above with luminosity. 

To find equilibria for different luminosities using our approach described above, we need to allow the functional form of our extension of either $L_{\mathrm{rad}}(r)$ or $\Phi(r)$ or both to change with luminosity. The behavior in the penetration region of $L_{\mathrm{rad}}(r)$ is likely to be the more important of these two. Our presently assumed functional form allows for a  gradual change in $L_{\mathrm{rad}}(r)$ that reflects the conditions seen in our higher-boost factor simulations. 

At lower luminosities our argument above indicates that the jump in entropy that results should be sharper, since it must begin at a lower value in order for the convection to reach this far. This conclusion is supported by our simulations. In our preliminary work to find equilibria for a series of luminosity values, we find that we can parameterize the jump in $L_{\mathrm{rad}}(r)$ from its value along a local adiabat and the ultimate value $L$, the total luminosity, in such a way that the thickness of this jump scales with the $1/3$ power of the luminosity. This power is conistent with and motivated by the discussion at end of \Sect{acc}. 

We can use a function like $\tanh(\mathrm{arcsinh}(x))$  with a jump thickness $\delta x$, over which $x$ increases from $-1$ to $+1$, that scales with $b^{1/3}$. We then find the radius, $r_\mathrm{CB}$, of the CB iteratively by demanding that the convective entropy flux must vanish there. To do that, we use the entropy equation, with our $\Phi(r)$ from the 3-D simulation, to solve for the radial gradient of the convective entropy flux, as described in this section.

A series of five such projected equilibrium states for the stellar model studied here, at the beginning of the main sequence when there is no composition gradient, and for boost factors from $b = 10^4$ to nominal are shown in \Fig{fig:boost-factor-dependent-transition-thickness}. These involve jump thicknesses of $151$, $70$, $32.5$, $15.1$, and \unit{7}{Mm} and result in jump-center radii of $1486$, $1572$, $1603$, $1616$, and $1622$ Mm.  These equilibrium models are presented here only to show the qualitative, rather than any quantitative implications of the 1-D convection zone modeling method presented in this section. If we define the CB as we have done here so far, then as the
luminosity increases, so also does the CB radius.  However, if we
define it as the center of the transition region, then its radius
decreases with increasing luminosity boost for models with
the same central entropy. We stress that the plots of the convective entropy fluxes in \Fig{fig:boost-factor-dependent-transition-thickness} are normalized by their luminosities.  Their appearance is counter-intuitive, because the center of the transition region at the CB moves slightly outward with decreasing luminosity.  However, the amplitude of this extension of the convective flux in the penetration region is $10^4$ times smaller than that shown for the most luminous case plotted in \Fig{fig:boost-factor-dependent-transition-thickness}.

In any case, the dependence of the equilibrium penetration depth, $r_\mathrm{CB} - r_\mathrm{SB}$, on the luminosity is quite weak, as we might have expected, in qualitative agreement with the results of \cite{Andrassy:24}. We see a change from a penetration depth, measured using the center of the transition region, of $166$ to \unit{302}{\Mm} as the luminosity varies over four orders of magnitude.  In this sense, even though the change in penetration depth is by a factor of two, it is nevertheless weak relative to the change in luminosity.

%%%%%%%%%%%%%%%%%%%%%%%%%%%%%%%%%%%%%%%%%%%%%%%%%%%%%%%%%%%%%%%%%%%%%
\section{Conclusions and Discussion}
\lSect{conclusions}

We have carried out an extensive study of core convection in a model star of 25 M$_\odot$ near the beginning of its main sequence life. We have focused our attention on the process of convective boundary mixing, by which the convection zone increases in size as the convective boundary, the CB, moves outward. We have simulated the convection using the PPMstar code, which employs accurate explicit numerical techniques for the gas dynamics and still more accurate moment-conserving advection techniques to track the concentration of the gas that is originally located above the convection zone. We find that despite the small value of the radiative diffusivity in the star, it is nevertheless essential to incorporate radiation diffusion into such simulations. Doing so allows the simulation to ultimately reach a dynamic and thermal equilibrium state in which radiation diffusion carries heat outward through the convective boundary, so that the convection zone does not need to continually expand in time. There is a relatively short timescale of several turnovers of the largest convective eddies in which the turbulent convection becomes thoroughly established and a dynamical equilibrium is achieved. That equilibrium can be expressed via the kinetic energy equation in 1-D, our \Eq{kin-eqn}. After this short time, even on a modest grid of only $768^3$ cells, the kinetic energy equation can be used together with radial profiles from the simulation to determine the kinetic energy dissipation rate $\Phi(r)$. 

We carry out such a 3-D simulation to measure $\Phi(r)$ using a luminosity boosted by a factor $b$ of 1000 or 10000, with radiative diffusion boosted by the same factor, to keep computation costs down.  So long as this boost does not significantly alter the near adiabatic structure of the convection zone up to the Schwarzschild boundary (SB), we have argued that $\Phi(r)$ will scale linearly with the boost factor, so that we can obtain this dissipation rate at the nominal luminosity of the star by a simple division by $b$.  $\Phi(r)$ is a key ingredient in our procedure for obtaining a 1-D model of the convective penetration region beyond the SB.

Boost factors, $b$, are widely used in simulating stellar core convection.  We have shown that the mass ingestion rate, and hence the rate of expansion of the core convection zone, scales linearly with $b$ (\Fig{entrainment}) and that the velocities in the convection zone scale with $b^{1/3}$ (\Fig{u-L}).  These scaling laws have been known for many years.  The first speeds up the approach to an equilibrium convection zone size, and the second speeds up to approach to the dynamical equilibrium inside the convection zone.  Thermal adjustments also are accelerated by the factor $b$.  Due to the lengths of these adjustment times and the cost of simulating the entire convection zone in 3-D, which is required to capture the largest and most important convective eddies and thus to evaluate $\Phi(r)$ correctly, boosting the luminosity is a technique that is used regardless of whether or not simulation codes are explicit, like ours, anelastic, or fully implicit.  We have found in our study here that the values of $b$ needed to bring the thermal adjustment time scale into a practical range also boost the velocities sufficiently to make our explicit approach practical as well.  We of course are simulating different stars on different grids, so that direct comparisons are difficult, but our luminosity boost factors are modest in comparison with much other work.

\begin{figure}
  \includegraphics[width=\columnwidth]{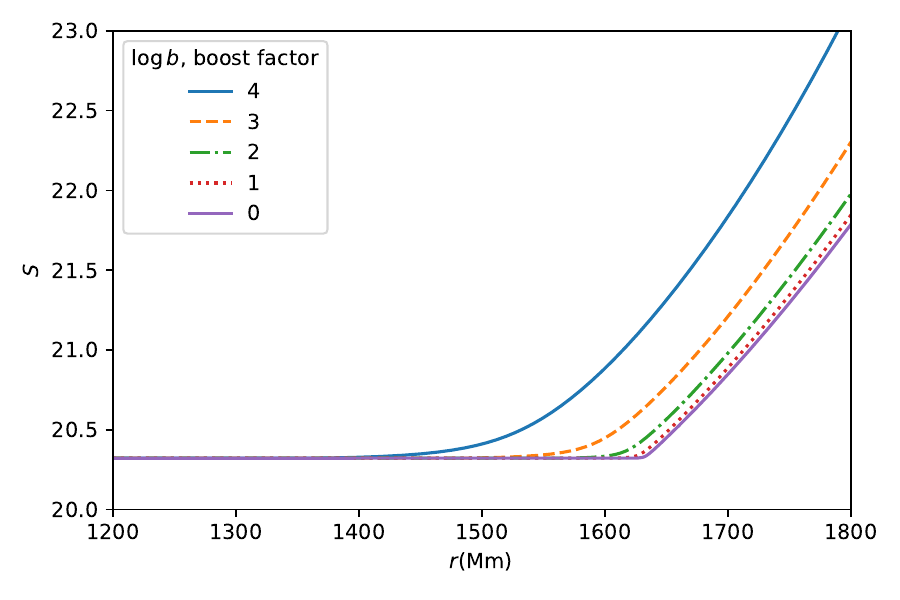}
  \includegraphics[width=\columnwidth]{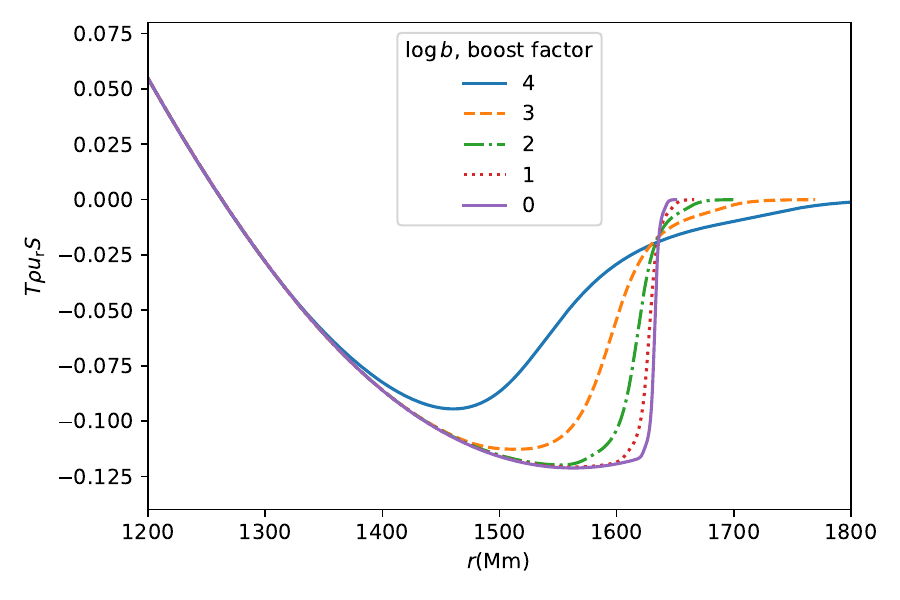}
  \caption{The radial profiles of entropy (top) and of the normalized entropy flux (bottom) are shown in the region of penetrative convection beyond the SB (at \unit{1316}{\Mm}) for a constant-$\mu$ stellar model of $\unit{25}{\Msun}$ (the same stellar model used for run M252, but at the beginning of its main sequence life).  A sequence of equilibrium models is shown that have been constructed by the method described in \Sect{penetration} for luminosity boost factors as indicated in the legend. These are corresponding to transition thickness parameters $\delta = 150.8$, $70$, $32.49$, $15.08$, and $\unit{7}{\Mm}$.  For the largest luminosity boost of 10000, the transition region extends nearly to the SB.  For the nominal luminosity of this star, the transition region in which the convection zone structure is not adiabatic is nearly a discontinuity, at a thickness of only \unit{7}{\Mm}.}
  \lFig{fig:boost-factor-dependent-transition-thickness}
\end{figure}
\begin{figure*}
       \includegraphics[width=\linewidth]{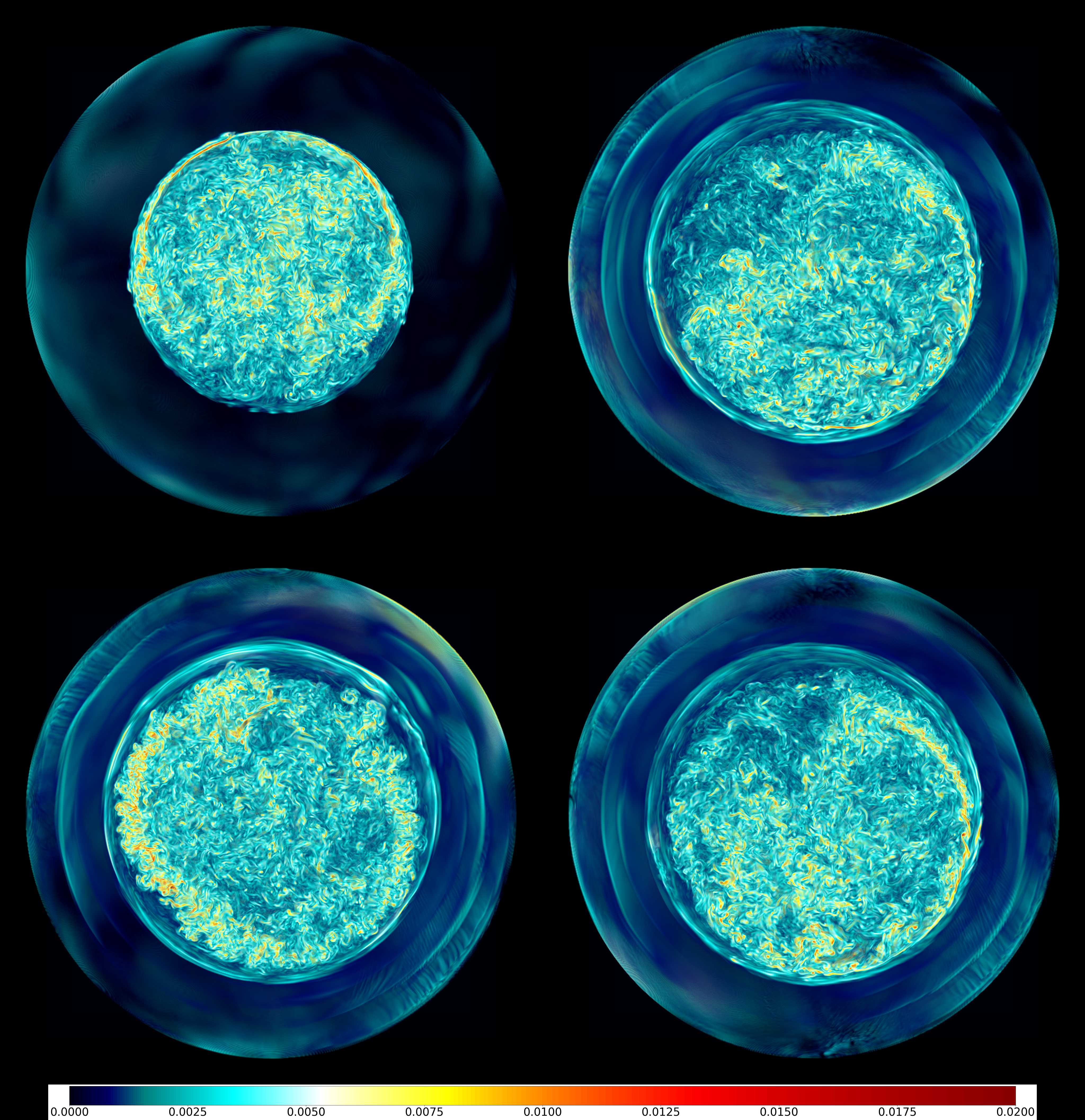}
       \caption{Four views of the vorticity magnitude in the far hemisphere of very long simulation M252 ($b=10000$, $896^3$ grid).  Top left: $t= \unit{12.06}{d}$, the dipole circulation pattern characteristic of core convection has become well established, and the radius of the CB is \unit{1535}{\Mm}. The other three views (clockwise from top right are at times $732.41$, $732.76$, and \unit{737.66}{d}) show later times, when the flow has developed a much larger region of penetrative convection above the SB.  The CB radii in these images are $1810$, $1811$, and \unit{1812}{\Mm}, respectively.}
    \lFig{M252image}
\end{figure*}

We have shown results of varying the luminosity boost factor $b$ over more than an order of magnitude.  The scaling behavior is observed as just discussed, and the resulting convective fluid behaviors are essentially the same at all boost factors studied. This result should not be a surprise, because at our largest boost factor of 10000 the Mach numbers in the convection zone are still small, having been boosted by only a factor of 21.5, so that the character of the convective flow is essentially unchanged.  This flow is visualized in \Fig{M252image}, where we show the magnitude of the vorticity at early and late times in our very long run M252.  

In the early flow, we see that the classic core-convection dipole circulation hugs the CB closely over about a quarter of the extent of this circle.  The flow separates from the CB where the prominent shear layers, marked by very strong vorticity (shaded yellow), bend inward from the boundary.  At top right, we see the flow much, much later.  The convection zone has expanded substantially, and the dipole circulation "contacts" the CB only along a very small segment, from which it immediately separates.  Just 0.35 days later, at bottom right, the dipole circulation has left the CB entirely, leaving a thin layer of somewhat higher entropy gas between it and the boundary.  In the image at the bottom left, despite the vigor of the dipole circulation flow, we see no contact with the CB, but we do see at about 2 o'clock, a strong gravity wave interfacial mode propagating along the CB, with a node in its flow pattern right at the CB radius. Our model of the convection zone identifies the thin, higher-entropy layer of convection zone gas right next to the CB as a key feature of this near-equilibrium penetrative convection structure.  This layer is generated by local heating from a declining radiative diffusion heat flux that is approaching the total luminosity in this region from above.
This flow can be compared with that shown in the lower-right panel of \Fig{image}, which was computed at triple the grid resolution for the same stellar model with a boost factor ten times lower.  

We have presented a 1-D method for finding the  equilibrium state of the convection zone complete with its penetration region. We first simulate the core convection in 3-D on a modest grid for several turnovers of the largest convective eddies, in order to establish a dynamical equilibrium. Using the 1-D kinetic energy equation, we solve for the kinetic energy dissipation rate, $\Phi(r)$, in the region inside the SB. This dissipation rate is an inherently 3-D phenomenon, because it depends upon the 3-D turbulent cascade, and it is affected by pressure accelerations in non-radial dimensions that do not cancel out upon averaging. To avoid this 3-D simulation, one might instead interpolate $\Phi(r)$ between such 3-D results obtained for similar stellar models. One might also use a mixing-length type model to obtain $\Phi(r)$.  It is very computationally costly to simulate the core convection flow until it comes into thermal as well as dynamical equilibrium. However, we can estimate the  equilibrium state quite accurately by analytically continuing $\Phi(r)$ and $L_{\mathrm{rad}}(r)$ between the SB and the CB, as described in \Sect{penetration}. 

The task ahead is to determine extrapolation functional forms for $L_\mathrm{rad}(r)$ and/or $\Phi(r)$ and their dependence on the boost factor that can be validated by simulations. This will inevitably involve an iterative procedure in which 1-D stellar evolution simulations that include our 1-D equilibrium prediction and 3-D simulations are alternated a few times until within the possible numerical accuracy dynamic and thermal equilibrium can be confirmed. The preliminary results of our 1-D model predictions as a function of boost factor shown in \Fig{fig:boost-factor-dependent-transition-thickness} imply that the transition layer is extremely thin at nominal luminosity. It would likely be best simulated in 1-D stellar evolution as an adiabatic step penetration layer, the thickness of which can be determined by our 1-D model calibrated with turbulent dissipation from 3D simulations. For our preliminary 1-D model prediction for nominal heating, shown in   \Fig{fig:boost-factor-dependent-transition-thickness}, which is as mentioned calibrated with a zero-age main-sequence simulation, the predicted penetrative overshoot would be $0.7H_\mathrm{p}$, which is almost a factor three larger than the prediction by \citet{Johnston:24} for the same mass.  

Our very long simulation, M252, has shed light on the approach to equilibrium in core convection.  Our analysis of the entropy equation reveals that at the SB the convective entropy flux does not vanish (see for example   \Fig{excessSfluxTur}), and therefore there must be a region of penetrative convection.  The convection is brought to an end by the joint actions of kinetic energy dissipation and a positive entropy gradient that develops of necessity in the penetration region as a result of the decrease with increasing radius of the radiative energy flux there. For the energy balance, especially in the penetration zone, the work done by global 3-D pressure fields is a key factor that enters our 1-D model predictions through the determination of the implied dissipation. Taking this effect into account is facilitated by our simulations adopting the correct 3-D $4\pi$ geometry that captures these global pressure fields as a result of the global dipole circulation. In the equilibrium state, the core convection maintains this propensity to organize into a prominent dipole circulation, with prominent shear layers where the diverted upward flow streams along the convective boundary.  However, in this equilibrium, as distinct from at earlier times when the CB is still moving outward, these shear layers separate the upwelling and diverted flow from a thin layer of heated gas that shares the well-mixed composition of the convection zone as a whole.  We have remarked upon this earlier, but it can best be seen in flow visualizations like those in \Fig{M252image}.  

\begin{acknowledgments}
The research reported here was supported by NSF through CDS\&E grants 1814181 and 2309101, travel grant 2032010, and through grants of access to the Frontera
computing system at TACC in Austin, Texas, where the bulk of the simulations were carried out and image rendering performed.  Partial support
was also provided by NSF through the JINA-CEE physics frontier center, award PHY-1430152.  Herwig acknowledges funding through an NSERC Discovery
Grant and a grant of access to the SciNet Niagara supercomputer operated by SciNet at the University of Toronto. Herwig also acknowledges
support for data analysis on the Astrohub online virtual research environment (https://astrohub.uvic.ca) developed and operated by the
Computational Stellar Astrophysics group (https://csa.phys.uvic.ca) at the University of Victoria and hosted on the Digital Alliance
Arbutus Cloud at the University of Victoria.  Woodward acknowledges support for local data storage and analysis from the Minnesota Supercomputing Institute.
\end{acknowledgments}

% \bibliography{HcoreM25radApJ}{}
% \bibliographystyle{aasjournal}

\end{document}